# Photoionization of the Be isoelectronic sequence: total cross sections

W.-C. Chu[1], H.-L. Zhou[1], A. Hibbert[2], and S. T. Manson[1]

[1]*Department of Physics and Astronomy, Georgia State University, Atlanta, Georgia 30303, USA*

[2]*School of Mathematics and Physics, Queen's University of Belfast, Belfast BT7 1NN, UK*

## Abstract

The photoionization of the four-electron beryllium-like isoelectronic series from the neutral to $Fe^{+22}$ has been studied for ground $^1S$ and metastable $^3P$ initial states. The wave functions of the final-state (target) ions were built using the CIV3 code. Both nonrelativistic *LS*-coupling *R*-matrix and relativistic Breit-Pauli (BP) *R*-matrix methods were used to calculate the cross sections in the photon-energy range between the first ionization threshold and the $1s^2 4f_{7/2}$ threshold for each ion. Our total cross sections compare well with experiment which is available for Be, $B^+$, $C^{+2}$, $N^{+3}$, and $O^{+4}$. The agreement between the present work and previous calculations is discussed in detail. The importance of relativistic effects is seen by the comparison between the *LS* and the BP results.

# I. INTRODUCTION

Since the universe is composed primarily of ions, the study of ions is of significance in astrophysics. Photoionization and the inverse process, electron-ion recombination, are important for the investigation of astrophysical, and other, plasmas. The Be-like four-electron closed-shell systems are particularly stable, which adds to their astrophysical importance. From the point of view of basic physics, the photoionization of ions is of interest as a fundamental process of nature. Furthermore, calculational technology has advanced to the point that it is possible to do extremely accurate calculations for systems with only four electrons. In addition, advances in target preparation and light source technology have led to a number of recent measurements of the photoionization of this isoelectronic sequence. Thus, the theoretical study of the photoionization of the Be-like isoelectronic sequence is rather timely as an adjunct to experiment and to assess the physics of the process.

The multiconfiguration Hartree-Fock (MCHF) method [1] and its relativistic counterpart, the multiconfiguration Dirac-Fock (MCDF) method [2], are used extensively in calculating discrete wave functions and discrete spectral properties of atomic systems. While MCHF and MCDF work with numerical orbitals, a similar method embodied in the CIV3 [3] code obtains equivalent results using expansions of analytical orbitals with adjustable parameters. These are the primary methodologies employed to calculate the wave functions of the discrete initial state and final residual ionic state in the calculation of the photoionization of atomic systems. To obtain the final-state continuum wave function, and the dipole matrix elements and the photoionization cross sections, state-of-the-art methods include the random-phase approximation with exchange (RPAE) [4,5], the relativistic random phase approximation (RRPA) [6], many-body perturbation theory (MBPT) [7], multi-channel quantum defect theory (MQDT) [8], and the *R*-matrix method [9,10]. For astrophysical modeling, however, data from the Opacity Project [11] and the Iron Project [12], which are based the *R*-matrix method, nonrelativistic and relativistic respectively, provide a majority of the data for light atoms and ions, and most of this data results from nonrelativistic Opacity Project calculations.

On the experimental side, the past decade has seen an explosion of absolute cross section measurements of photoionization owing to the availability of intense light sources such as the Advanced Light Source (ALS) in USA, ASTRID in Denmark, the Photon Factory in Japan, and SuperACO in France. The merged-beam technique has been used in most measurements since the intensity of third-generation light sources has made this method practical [13].

In the present work, photoionization cross sections of 14 members of the Be isoelectronic are calculated using the relativistic Breit-Pauli $R$-matrix (BPRM) method [14]. In particular, the systems included in this investigation are Be, $B^+$, $C^{+2}$, $N^{+3}$, $O^{+4}$, $Ne^{+6}$, $Mg^{+8}$, $Si^{+10}$, $S^{+12}$, $Ar^{+14}$, $Ca^{+16}$, $Ti^{+18}$, $Cr^{+20}$, and $Fe^{+22}$. We report the ground state and metastable state cross sections and comment on their general behavior along the sequence. The study also provides data for astrophysical models which is both complete and whose accuracy is assessable *via* comparison with experiments and other calculations. Extant experimental measurements include Be [15, 16], $B^+$ [17], $C^{+2}$ [18], $N^{+3}$ [19, 20], and $O^{+4}$ [21]. Previous calculations include, in addition to Opacity Project [11, 22], work on Be [23, 24], $B^+$ [25, 26], $C^{+2}$ [27-29], $N^{+3}$ [30], and $O^{+4}$ [31]. Note that in the experimental results for the various ions, the target ion beams are often mixtures of the ground state $1s^22s^2\ ^1S_0^e$ and metastable states $1s^22s2p\ ^3P_{0,1,2}^o$. Thus, in order to compare with experiment, calculations of both ground and metastable states are required, and the resulting cross sections are combined to match the composition of the experimental beams. In practice, however, the experimental fractions are often unknown and they are obtained by varying the fractions of the calculated cross sections to obtain optimal agreement with experiment. In addition, calculations using the nonrelativistic (*LS*-coupling) $R$-matrix method [9,10], in each case, are also performed in an effort to assess the importance of relativistic interactions along the sequence.

In Sec. II, a brief description of $R$-matrix theory and the details of the calculation are presented. Sec. III reports the total cross sections resulting from our calculation, along with available experiments and other calculations. Comparisons and comments are also given here. Sec. IV presents the conclusions.

## II. THEORY AND METHOD OF CALCULATION

The nonrelativistic (*LS*-coupling) photoionization processes given by

$$1s^22s^2(^1S^e) + h\nu \rightarrow [1s^2nln'l' + e^-(kl'')](^1P^o) \tag{1}$$

for ground state photoionization and by

$$1s^2 2s 2p(^3P^o) + h\nu \rightarrow [1s^2 nln'l' + e^-(kl'')](^3S^e, ^3P^e, ^3D^e) \tag{2}$$

for metastable state photoionization are considered in this work. In the relativistic case, the corresponding transitions are given by

$$1s^2 2s^2(^1S_0^e) + h\nu \rightarrow [1s^2 nln'l' + e^-(kl'')](^1P_1^o) \tag{3}$$

and

$$1s^2 2s 2p(^3P_{0,1,2}^o) + h\nu \rightarrow [1s^2 nln'l' + e^-(kl'')](^3S_1^e, ^3P_{0,1,2}^e, ^3D_{1,2,3}^e) \tag{4}$$

in *LSJ* terms. However, in our Breit-Pauli calculations, the selection rule specifies only $0^e \rightarrow 1^o$ for ground state and $0^o \rightarrow 1^e$, $1^o \rightarrow (0^e, 1^e, 2^e)$, and $2^o \rightarrow (1^e, 2^e, 3^e)$ for metastable state transitions, where all these $J\pi$ symmetries contain the contributions from all the possible *LS* terms. To construct the wave functions, we optimize the Slater-type orbitals, $o_{nlm_l}$, first discussed in [33]. We take the 1*s* and 2*s* orbitals from HF calculation by Weiss [34] and by Clementi and Roetti [35] and optimize other orbitals up to 4*f* with the CIV3 code [3]. In *R*-matrix theory [9, 10], the system consists of *N*+1 electrons; the final state is the *N*-electron final state of the ion (known as a target state for historical reasons) plus a photoelectron. The *N*-electron configurations $\phi_i$ are antisymmetric combinations of the one-electron orbitals. Using these configurations as basis set, we diagonalize the *N*-electron Hamiltonian $H^{(N)}$ for each *LS* term to get target states, $\Phi_j$, the final states of the ion minus the photoelectron. Of course, in an exact calculation, there must be an infinite number of configurations to complete the basis set, but we include only the ones that we believe to be the most important from a physical point of view. In the present work, 9 configurations are employed: $1s^2 2s$, $1s^2 2p$, $1s^2 3s$, $1s^2 3p$, $1s^2 3d$, $1s^2 4s$, $1s^2 4p$, $1s^2 4d$, and $1s^2 4f$. In the relativistic (BP) calculation, relativistic corrections are added to the Hamiltonian and the expansion of target states is specified by *LSJ* terms. The resulting target state energies, shown in Table I for four of our Li-like target (final state) ions, are compared with NIST values [36]. The agreement, an indication of the accuracy of the N-electron wave functions, is seen to be excellent.

*R*-matrix theory divides configuration space into an inner and an outer region, separated by a spherical shell of radius $r_a$ centered at the nucleus. In the inner region, all *N*+1 electrons are treated on an equal footing and all exchange effects are considered. The (*N*+1)-electron wave function for the symmetry $SL\pi$ (and *J* in the BP calculation) is given by

$$\Psi_k^{SL\pi} = \mathcal{A} \sum_{ij} c_{ijk} \overline{\Phi}_i(S_i L_i; x_1, x_2, \ldots, x_N, \hat{x}_{N+1}) \frac{F_j(r_{N+1})}{r_{N+1}} + \sum_j d_{jk} \chi_j^{SL\pi} \quad (5)$$

where the $\overline{\Phi}_i$ are target state wave functions coupled with the angular and spin part of the photoelectron, $F_j$ are the continuum wave functions of the photoelectron, and $\chi_j$ are the (N+1)-electron bound state wave functions. The summation over *ij* of the first term is over all open channels of $\Phi_i$ and over all continuum electron orbitals $F_j$ to give *SL*π symmetry. The summation over *j* of the second term is over all (N+1)-electron bound states to ensure the completeness of the basis. In the outer region, the system is simplified as a two-body system where the photoelectron is in a long range multipole field. The wave function in this region is given by

$$\Psi_k^{SL\pi} = \sum_{ij} c_{ijk} \overline{\Phi}_i(S_i L_i; x_1, x_2, \ldots, x_N, \hat{x}_{N+1}) \frac{F_j(r_{N+1})}{r_{N+1}}. \quad (6)$$

Notice that there is no exchange term between the photoelectron and any other electron in this outer region. Then enforcing continuity at the boundary, we get the wave functions in both regions with the *R*-matrix as a connection between the parts. The Breit-Pauli version includes the relativistic corrections to the Hamiltonian to improve the accuracy in dealing with heavier ions.

When the initial wave function and the final wave function (for some specific total energy) are known, the photoionization cross section *σ* is calculated by using the electric dipole approximation to evaluate the transition matrix.

Both initial and final states use the same set of orbitals optimized in CIV3. Since these orbitals are optimized for the target states, this might cause some problem in constructing the (N+1)-electron initial state. As seen in Table II, where ground initial state binding energies of all ions in our calculations are compared with NIST values [36], this is only a problem (and a slight one) at the very high end of the sequence, and similarly in Table III for the metastable initial state binding energies. Increasing the basis set to take care of this problem simply shifts the cross section, and does almost nothing else, so it was not considered to be worthwhile.

## III. RESULTS AND DISCUSSION

### A. Cross Sections

The total photoionization cross sections, calculated using the BPRM, along to Be-like isoelectronic sequence for the $^1S_0^e$ ground state are shown in Fig. 1 and for the $^3P^o$ metastable state in Fig. 2 over photon energy range (in Rydbergs) from the lowest ionization threshold up to 4$f$ threshold. Actually, there are three metastable levels, $^3P_0^o$, $^3P_1^o$ and $^3P_2^o$, and their cross sections are almost identical. Thus, rather than showing each of them, Fig. 2 shows a statistical average of the three cross sections. Since the agreement of length and velocity gauges in our calculations is excellent (within a few percent, at worst), all of our calculated cross sections are shown only in length gauge; it is of importance to emphasize that agreement between length and velocity is a necessary condition for accuracy of calculated photoionization cross sections. Some general features of the evolution of these ground and metastable state cross sections are noted.

The resonances dominating the threshold region of each of the cross sections are the 2$pns$ and 2$pnd$ autoionizing states. At higher energies are the manifolds of resonances converging to the $n = 3$ and the $n = 4$ states of the final state ion. It is interesting to note that the cross sections for the ground and metastable states are qualitatively similar, despite the fact that they are of differing parities, thereby connecting to opposite parity continua, and of different spin-multiplicity, so they connect to continua of different spin-multiplicity.

The area under the cross section curves, in Mb-Ryd, divided by 8.07 is the total oscillator strength in over that energy region [37]; owing to the well-known sum rule [37], the total oscillator strength from the outer shell is 2, the number of outer-shell electrons. This sum includes the discrete oscillator strengths for the excitations below the ionization threshold, and these are large, e.g., in neutral Be, the 2$s$→2$p$ transition has an oscillator strength of 1.36 [38]. It is found, from the present results, that the oscillator strength from the first threshold to the 1$s$4$f_{7/2}$ threshold, is approximately 0.4 and is about the same for all of the $Z$ considered and for both initial states. Then, since the energy scale increases roughly as $Z^2$ (the hydrogenic energy scaling), it is evident that the cross sections must decrease as $1/Z^2$ to preserve the oscillator strength; this is exactly what is seen in Figs. 1 and 2.

The energy separations between *nl* and *nl'*, on the other hand, do not increase as $Z^2$ along the isoelectronic sequence; they increase, but only roughly as *Z*. For example, the interval between 2*s* and 2*p* thresholds is 0.293 Ryd in Be (*Z*=4), which is about 27% of the range from 2*s* threshold to 4*f* threshold. This percentage is down to 8.8% in Ne$^{+6}$ (*Z*=10), 5.5% in S$^{+12}$, and 4.1% in Fe$^{+22}$. Of course in the $Z \to \infty$, hydrogenic, limit levels of the same principal quantum number are degenerate (nonrelativistically) asymptotically, so a different energy dependence is expected. In any case, owing to the differences in the dependences of the thresholds of differing *n* and *l* along the isoelectronic sequence, the overlapping of resonances converging to threshold having the same principal quantum is considerably altered as a function of *Z*. Of course, with increasing *Z*, spin-orbit effects, which increase as $Z^4$ become important, so the situation for the higher-*Z* ions is rather more complicated.

The width of a resonance indicates the strength of the Coulomb matrix element of the quasi-discrete resonance state with the final continuum state. The resonance widths increase slowly with *Z* while the energy range grows as $Z^2$ as described above; actually, in the hydrogenic limit, the resonance widths are independent of *Z* [39]. As a total effect, the widths of the resonances relative to the energy range decrease with increasing *Z*, making $\sigma(E)$ a smoother function and the resonance structures less important in the sense that less of the energy range is resonant and more is nonresonant background cross section. Thus, in the heavier ions, the background cross section can be fit more easily by a simple function without too much disturbance by the resonances. At the lower end of the sequence, owing to the extent of the resonance widths, this is more problematic. Furthermore, the widths also decrease relative to the energy separation of the resonances, with increasing *Z*, strongly affecting the resulting cross section in the resonance region.

There is clearly a wealth of information concerning the evolution of the resonances along the isoelectronic sequence, but the details will be presented in a separate paper.

### B. Comparison with Experiment`

#### 1. Be

For the neutral Be atom two experiments in separated energy ranges have been reported, both performed at the University of Wisconsin Synchrotron Radiation Center (SRC). Wehlitz *et* *al* [15] measured the ground state cross section from the ionization threshold at 9.3227 eV to the 2*p* threshold at 13.277 eV with energy step ($\Delta E$) 20 meV below, 5 meV beyond, 12.60 eV and monochomator bandpass of 12 meV, which we take as the full width at half maximum (FWHM) when convoluting our theoretical cross section calculated with energy step $\Delta E$ = 68 μeV. Fig. 3 shows the present BPRM and nonrelativistic *LS R*-matrix cross sections along with the experimental results from 9.2 eV to 13.3 eV. From the comparison of the BPRM and *LS* results it is clear that relativistic effects in the photoionization of neutral beryllium are negligible. The present theoretical results show excellent agreement with experiment below about 12.5 eV, but the experimental peaks seem to be truncated, compared to theory, at higher energy. It is evident, however, that the positions of the resonances are in excellent agreement over the whole range. We can think of no explanation for the lower resonances in the series being more accurate than the higher members, so this could be an experimental problem.

In the higher energy range, near 3*s* and 3*p* thresholds, SRC measurements were made by Olalde-Velasco *et al* [16] with energy step 15 meV from 16 eV to 19.5 eV and 5 meV beyond 19.5 eV. With our energy step $\Delta E$ = 68 μeV, we convolute our result with FWHM = 27.5 meV and FWHM = 7.5 meV, below and beyond 19.5 eV respectively to compare with experiment. BPRM and nonrelativistic cross sections as well as the measurement are shown in Fig. 4. Just as in the lower energy range, the difference between BPRM and nonrelativistic results is negligible. The overall background cross section in the calculation is about 0.6 Mb higher than the measurement, and there is a 0.1 eV energy shift between calculation and measurement.

## 2.  $B^+$

The $B^+$ calculation is compared with the measurement by Schippers *et al* [17] at Advanced Light Source (ALS) at Lawrence Berkeley National Laboratory (LBNL). The measurement was done with $\Delta E$ = 4 meV from 22.50 eV to 31.26 eV, whereas our calculation was performed with $\Delta E$ = 13.6 μeV. In the calculation, we assume that the initial beam has 71% $^1S_0^e$ ground state ions and 29% $^3P_1^o$ metastable state ions [17]. The calculations and the measurement are shown in Fig. 5. Both calculations are convoluted with FWHM = 25 meV. Below the ground state ionization threshold at 25.091 eV (calculated result), the cross section is generated from metastable state photoionization only. The theoretical threshold for the metastable state is 20.44 eV in the present calculation, which the experiment could not identify because the photon flux was too low. The difference between the BPRM result and the nonrelativistic result is clearly due to the splitting of the resonances in this region. BPRM clearly shows the peaks that are missing in the nonrelativistic cross section. The BPRM result matches the measurement well except for an overall energy shift of about 0.05 eV.

### 3. $C^{+2}$

In the $C^{+2}$ ion, the experiment was conducted by Müller *et al* [18] at ALS. It was done with $\Delta E = 4$ meV from 40.84 eV to 56.98 eV. We assume that 60% of the ions were in the $^1S_0^e$ ground state and 40% in the $^3P^o$ metastable state in the initial ion beam; specifically 30% $^3P_0^o$ and 5% each of $^3P_1^o$ and $^3P_2^o$ [18]. The calculations used an energy step size $\Delta E = 12.2$ μeV and are convoluted with FWHM = 30 meV. As shown in Fig. 6, similar to the $B^+$ case, the splitting is the biggest difference between the two calculations. Compared with the experiment, the experimental threshold energies 41.39 eV and 47.89 eV are higher than the present values 41.28 eV and 47.81 eV for the metastable state and the ground state respectively. The theoretical background cross section is a bit higher than experimental cross section near the $2p_{3/2}$ threshold of $^1S_0^e$ ground state at 55.8987 eV. Other than that, our BPRM result matches the experiment well in all resonance positions and widths.

### 4. $N^{+3}$

Experimental work on $N^{+3}$ ions was performed by Bizau *et al* [19] at ASTRID at the University of Aarhus. They obtained the cross section with $\Delta E = 100$ meV in the range 63.00 eV to 90.00 eV. In our calculation, we used $\Delta E = 13.6$ μeV and convoluted the result with FWHM = 230 meV. The fractions of $^1S_0^e$ ground state and $^3P^o$ metastable state are assumed to be 65% and 35% respectively [19]; in the absence of any more detailed information on the excited initial states, we assumed that the three metastable states were populated statistically. In Fig. 7, it is seen that the difference between BPRM and the nonrelativistic results is that the peak heights and strengths in the metastable region below 79 eV are larger in the BPRM case. This is likely because the inclusion of relativistic effects in the BPRM calculation opens photoionization channels that are forbidden in the nonrelativistic *LS* case, thereby increasing the resonance oscillator strengths. Comparing our BPRM calculation with experiment, we find that in the low energy range where only the metastable state contributes, the background agrees well but the resonances are slightly weaker than experiment, but much closer than the nonrelativistic results. In the higher energy range where ground state photoionization dominates, the experimental cross section is very noisy and it is difficult to pick out the higher resonances, but the first few show reasonable agreement. The nonresonant background cross sections are in good agreement in the lower energy region where only the metastables contribute, but theory is a bit higher than experiment at the higher energies where ground state photoionization dominates. In addition, there has been some recent high-resolution experimental work in very narrow energy ranges reported [20]: the region of the metastable thresholds, and the region of the $2p5p$ resonances. Apart from a small energy shift, our calculations, convoluted with the experimental resolution (not shown), show excellent agreement.

## 5.  $O^{+4}$

The measured cross section was obtained by Champeaux *et al* [21] at SuperACO at LURE in France with $\Delta E = 56.4$ meV in the range from 99.60 eV to 129.75 eV. In the calculation, we had an energy step size $\Delta E = 13.6$ μeV and it was convoluted with the experimental FWHM = 250 meV. The experiment reported fractions of 50% $^1S_0^e$ ground state ions and 50% $^3P^o$ metastable state ions in the beam; since no breakdown of the metastable part of the beam was reported, we assumed a statistical distribution as in the $N^{+3}$ case, discussed above. Similar to $N^{+3}$ ions, the BPRM cross section shows stronger resonances than those in the nonrelativistic one in the region of the spectrum due to metastable photoionization only. The reason is presumably the same as that in the $N^{+3}$ case. The comparison between our calculation and the experiment is also similar to the $N^{+3}$ case. The background and resonance positions match well with experiment in the metastable region, but it is much harder to read the resonance information in the ground state region in the experiment.

### C.  Comparison with Other Calculations

The details of Opacity Project (OP) are described by Seaton [11]; the photoionization of Be-like ions was studied by Tully *et al* [22]. OP includes atomic data of 15 isoelectronic ions up to $Fe^{+22}$ based on the nonrelativistic *R*-matrix calculation. To give the flavor of the comparison of the OP results with the present BPRM data, and how it changes along the isoelectronic sequence, the comparison for Be, $Ne^{+6}$, $Ar^{+14}$ and $Fe^{+22}$ are shown in Figs. 9-12 respectively for both ground and metastable states; our BPRM results are the statistical average of the three $^3P^o$ metastable states. For Be, Fig. 9, the ground state comparison shows that the OP ground state threshold energy is a bit lower than the BPRM result and, thus, lower than the experimental (NIST) value, by about 0.1 eV. For the metastable state, the OP threshold is too low by considerably more than that. Consequently, the OP metastable cross section at threshold is about 10% too high. In addition, careful comparison reveals that the OP resonances are at somewhat different energies than the present BPRM results. Since the latter are in good agreement with experiment, as detailed above, it is evident that the OP calculation is lacking in this respect as well. Most important, however, is that the energy mesh used in the OP calculation is seen to be much too coarse to correctly reproduce the resonances in both ground and metastable states. This results is much of the resonance oscillator strength being absent from the OP cross sections, as seen in Fig. 9.

For $Ne^{+6}$, shown in Fig. 10, the comparison is qualitatively similar, but the discrepancies are quantitatively greater, owing to the fact that relativistic interactions are more important in $Ne^{+6}$ than in neutral Be; for $Ne^{+6}$, the OP thresholds are too low by several eV and the 2p thresholds are seen to be even worse, especially for the ground state. Further, owing to the energy step size, the higher 2*pnl* resonances are absent from the OP results.

Going up to $Ar^{+14}$, Fig. 11, the comparison is seen to be dramatically worse. The OP thresholds are off by of the order of 20 eV. In addition, the resonances are almost unobservable; and those that are seen are at rather incorrect energies. Only the OP background, nonresonant cross section is reasonably good in this case. However, we note that the OP background cross section is not significantly better than the results of a central-field Hartree-Slater (HS) calculation which give a threshold value of the ground state cross section of about 0.11 Mb [40], in good agreement with these results.

For $Fe^{+22}$, Fig. 12, the comparison is similar to the $Ar^{+14}$ case, but even further apart. The OP thresholds are so far off that the whole OP resonance region converging to 2*p* ranges from 137 Ryd to 142 Ryd, but it ranges from 143 Ryd to 148 Ryd in the present work, and there no overlap between these regions between the two calculations. The OP thresholds are off by ~ 100 eV! Again, the OP background cross section is reasonably accurate, but so is the simple HS result of the compilation of Ref. 40.

Several photoionization calculation of lower members of the isoelectronic sequence, Be [23, 24], $B^+$ [25, 26], and $C^{+2}$ [27, 28], have been calculated for both ground and metastable states using a variational $R$-matrix method (VRM) [41]. In these calculations, the $1s^2$ core is replaced by an effective potential which is optimized by comparison of binding energies with experiment, and the wave functions of the two outer electrons are solved by the Schrödinger equation. The basic differences between this method and the present calculation are: the variational $R$-matrix calculations are nonrelativistic while ours include relativistic effects; also, that method is semi-empirical, based on optimizing the potential due to inner-shell electrons to fit experimental energies, while ours is purely *ab initio*. In Figs. 13-15 a comparison of the present BPRM photoionization cross sections with the variational $R$-matrix results are shown for Be, $B^+$, and $C^{+2}$ ions, respectively; all cross sections presented are in length gauge since the length and velocity results essentially coincide in both calculations. Our cross sections are generally in good agreement with the variational $R$-matrix results, but there are a few differences in all three cases. First, both our $^1S_0^e$ ground state and $^3P^o$ metastable state ionization thresholds are lower than the variational $R$-matrix values. Second, the inclusion of relativistic effects opens more ionization channels, which cause splitting of some of the resonances, as seen in the Figures. Third, there are differences in the shapes of the resonances at the beginning of some Rydberg series. For example, in Fig. 13, the thin resonance near 11.8 eV is seen to have a different shape in the two calculations which amounts to almost a vertical flip, and a similar flip occurs around 23.2 eV in $B^+$. This means in the analysis of Beutler-Fano profile of resonance, the $q$ value has opposite sign in the two calculations, which implies either the discrete or the continuum final state at the corresponding energy has a phase difference between the calculations.

There have also been calculations of the photoionization of $C^{+2}$ [29,30], $N^{+3}$ [30] and $O^{+4}$ [31] using orbitals obtained with the SUPERSTRUCTURE code [42] and nonrelativistic $R$-matrix to calculate the cross sections, except for $C^{+2}$ where a relativistic calculation was also done. Figs. 16-18 show the comparison of our BPRM cross sections with the previous nonrelativistic results [30,31] for both ground and metastable states of $C^{+2}$, $N^{+3}$, and $O^{+4}$ ions, respectively. The general features of the cross sections, such as the ionization thresholds, resonance positions and widths match pretty well. As seen in these figures, however, the main difference between their calculations and ours is the splitting of resonances due to relativistic effects. Note further, that the comparison of their results with our nonrelativistic cross sections (not shown) show excellent agreement, thereby indicating that these earlier calculations include the important physics, except for the relativistic effects. To emphasize this point, note that using the same methods for discrete states, but with a BPRM formulation for the continuum states, the relativistic photoionization calculation for $C^{+2}$ was performed [29] and included both ground and metastable states [43]; the comparison with the present BPRM results are shown in Fig. 19, where excellent overall agreement is seen, both as to resonance positions and background nonresonant cross sections. Some small differences are seen in the amplitudes and shapes of the very narrow resonances. Also, for the ground state cross section, the peaks of the Ref. 29 higher resonances of the major series are erratic, while our results are not. We attribute this to a lack of sufficient density of energy points in the neighborhood of these resonance peaks in Ref. 29. A similar BPRM calculation [31] for $O^{+4}$ has been performed (not shown) and the agreement with the present calculation is similar to that of the C+2 comparison exhibited in Fig. 19.

BPRM calculations of the photoionization of $B^+$ [17], $C^{+2}$ [18], and $N^{+3}$ [20] using the same discrete orbital methodology as used in the present paper have also been reported. For the $B^+$ and $C^{+2}$ cases, the results for the experimental admixture of ground and metastable states, suitably convoluted with the experimental width, are shown in Figs. 5 and 6 respectively. As can be seen, they are almost identical to the present BPRM results. This is hardly surprising since the two calculations used essentially the same target states, although somewhat different versions of the BPRM code [44]. The very slight differences in the results around some of the narrow resonances can be largely traced to not using enough energy points in the energy mesh to completely characterize the resonance [44]. For the case of $N^{+3}$ [20], where many energy points are used in very narrow energy ranges, our calculated results match the previous result essentially exactly. In any case, the agreement strongly suggests that both calculations were done correctly.

### D. Comparison Between *LS*-coupling and BP Calculations

To pinpoint the influence of relativistic effects, calculations have been performed at both the *LS*-coupling and Breit-Pauli (BP) levels, using exactly the same radial basis set and radial wave functions; this procedure insures that any differences in the cross sections resulting from the two levels of calculation are due solely to relativistic effects. For the first five members of the sequence, both the *LS*-coupling and BP calculations are shown in Figs. 3-8. In this low-*Z* part of the isoelectronic sequence, it was seen that there were only small differences between *LS* and BP results; of importance, however, is that in every case, the relativistic result is closer to experiment. Owing to the experimental resolution that our theoretical results have been convoluted with, it is difficult from these Figures to make any statement about how the importance of relativistic effects changes with increasing *Z*. However, looking at our *un*convoluted results (not shown), it is clear that relativity becomes more important with increasing *Z*.

To explore this further, the comparison for Ne$^{+6}$ is shown in Fig. 20 with no convolution for both ground and metastable state cross sections; the metastable BP result presented is a statistical average of the cross sections of the three $^3P_j^o$ metastable states. While the background cross sections are the same, in both cases, the resonances are seen to differ in position size and shape, particularly for the excited state. As an example, for the ground state resonance at about 19.15 Ryd, the nonrelativistic position is about 0.04 Ryd (0.54 eV) lower than the BP location, which is caused by the relativistic shift of the ground state energy plus the shift of the threshold energies of the final states of the ion. Further, for the photoionization of the initial excited metastable states, there is a marked difference in the size and shape of the resonances between *LS* and BP results that is not evident for ground state photoionization. This is seen in Fig. 20 in the 21.2-Ryd photon energy region where a large narrow nonrelativistic resonance is "surrounded" by a number of smaller relativistic resonances. This disagreement occurs primarily because the three relativistic metastable states have differing threshold energies which results in the resonances being located at somewhat different energies. Thus, in a statistical average, instead of a single resonance, as in the *LS* case, there are three resonances "sharing" the oscillator strength which is more or less preserved.

As *Z* increases further, the energy shifts are expected to grow larger, and, looking at the comparison for S$^{+12}$, shown in Fig. 21, this is true. Here, shifts in thresholds and resonances of about 0.2 Ryd (2.7 eV) are evident. And, for Fe$^{+22}$, shown in Fig. 22, the shifts of thresholds and resonances are as large as 1.6 Ryd, more than 20 eV are noted. And, for both S$^{+12}$ and Fe$^{+22}$ the discrepancies of the resonances for the excited state are evident, just as in the Ne$^{+6}$ case discussed above. Furthermore, while only three of the higher members of the isoelectronic sequence are shown in detail, there is nothing special about those particular ions; the above discussion applies to all of the higher members of the sequence. In any case, it is clear that relativistic shifts become more and more important with increasing $Z$.

The scale of the plots of Figs. 20-22 is such that the splitting of resonances is not seen. However, they are certainly there, and the splitting increases with $Z$ also. The very small splitting of resonances for $B^+$ and $C^{+2}$, seen in Figs. 5 and 6 respectively, are of the order of 100 meV. At the other end of the scale, for $Fe^{+22}$, the splittings of the resonances can be as large as of the order of 10 eV. Also not evident from the plots is that the overlapping of resonance series is rather different in the *LS* and BP cases, at the higher $Z$'s. Thus, relativistic effects play an important role on perturbing the resonance positions, splitting them into doublets, and changing the overlaps among resonance series converging to different states of the final-state ion. The details of the resonances, and how they change as a function of $Z$, will be reported and analyzed in a separate paper.

## IV. CONCLUSIONS

From the comparison of our calculated cross sections for the ground and metastable states of a number of members of the Be isoelectronic sequence with experiment (where available) and previous calculations, we find that the relativistic Breit-Pauli *R*-matrix methodology, along with an extensive high-quality set of discrete orbitals to represent the final ionic (target) states and the initial states, provides an extremely accurate description of the photoionization process in the four-electron system. This is further confirmed by the excellent agreement between length and velocity gauges (within a few percent), and by the excellent agreement of the initial state binding energies and target state excitation energies with the NIST data compilation, presented in Tables I-III.

In our study, it has been found that the overall magnitudes of the cross sections decrease with $Z$, which is necessary to satisfy the oscillator strength sum rule [37]. Inclusion of relativistic effects is found to be of importance to achieve high accuracy even at the lowest values of $Z$, owing to the splittings and shifts of the resonances engendered by these relativistic effects. And, since relativistic effects in energies increase as $Z^4$, while electrostatic energies increase as $Z^2$, these effects become much more important at the higher *Z*'s; for $Fe^{+22}$, for example, relativistic resonance energy shifts of close to 100 eV were found. And for the excited states, relativistic effects were even more important because the single nonrelativistic $^3P^o$ state is split into three states, $^3P^o_0$, $^3P^o_1$ and $^3P^o_2$, each with a different threshold energy. Further, it is important to note that these conclusions should be quite general and not restricted just to the four-electron Be-like systems studied in detail here. Many of the cross sections were seen to exhibit large numbers of very narrow resonances. Thus, to calculate cross sections which include all of the oscillator strength, the calculational energy mesh must be dense enough to include the maxima of these resonances. Furthermore, as seen in some of the comparisons of theoretical cross sections, some of the reported results contain the proper physics but are still inaccurate owing to the failure to include enough points in the energy mesh.

The results presented in this paper concern only the gross characteristics of the total cross sections. Clearly there is a wealth of information on the partial cross sections, cross sections for producing the various states of the three-electron final state ion. In addition, there are many series of autoionizing resonances converging to the various $1s^2nl$ thresholds which have not yet been described or analyzed, including overlapping of series converging to states of different principal quantum number, and how the resonances and the overlaps evolve as a function of *Z*. These issues shall be dealt with in future publications.

## Acknowledgments

This work was supported by U.S. Department of Energy, Division of Chemical Sciences, and NASA. We are indebted to Alfred Müller, Stefan Schippers, Ron Phaneuf, R. Wehlitz and S. B. Whitfield for supplying us with numerical tables of their published data. We also acknowledge a number of helpful discussions with Brendan McLaughlin. All calculations were performed using National Energy Research Scientific Computing Center (NERSC) computational resources.

Table I. Energy levels (in Rydbergs) in the present work and in NIST data [34] for the Li-like target ions $Be^+$, $Ne^{+7}$, $S^{+13}$, and $Fe^{+23}$. All energies are relative to the $1s^2 2s\ ^2S^e_{1/2}$ ground state energy.

|  | Present | NIST | Present | NIST | Present | NIST | Present | NIST |
|---|---|---|---|---|---|---|---|---|
|  | $Be^+$ Z=4 | | $Ne^{+7}$ Z=10 | | $S^{+13}$ Z=16 | | $Fe^{+23}$ Z=26 | |
| $2p\ ^2P^o_{1/2}$ | 0.293 | 0.291 | 1.176 | 1.168 | 2.053 | 2.045 | 3.602 | 3.572 |
| $2p\ ^2P^o_{3/2}$ | 0.293 | 0.291 | 1.191 | 1.183 | 2.189 | 2.182 | 4.745 | 4.745 |
| $3s\ ^2S^e_{1/2}$ | 0.799 | 0.804 | 10.018 | 10.023 | 29.345 | 29.344 | 84.508 | 84.497 |
| $3p\ ^2P^o_{1/2}$ | 0.876 | 0.879 | 10.338 | 10.341 | 29.911 | 29.908 | 85.515 | 85.460 |
| $3p\ ^2P^o_{3/2}$ | 0.876 | 0.879 | 10.342 | 10.346 | 29.951 | 29.949 | 85.844 | 85.815 |
| $3d\ ^2D^e_{3/2}$ | 0.888 | 0.894 | 10.450 | 10.457 | 30.163 | 30.156 | 86.234 | 86.197 |
| $3d\ ^2D^e_{5/2}$ | 0.888 | 0.894 | 10.451 | 10.455 | 30.175 | 30.169 | 86.343 | 86.321 |
| $4s\ ^2S^e_{1/2}$ | 1.047 | 1.052 | 13.386 | 13.392 | 39.368 | 39.364 | 113.627 | 113.584 |
| $4p\ ^2P^o_{1/2}$ | 1.077 | 1.082 | 13.517 | 13.522 | 39.599 | 39.594 | 114.020 | 113.989 |
| $4p\ ^2P^o_{3/2}$ | 1.077 | 1.082 | 13.519 | 13.524 | 39.616 | 39.612 | 114.156 | 114.135 |
| $4d\ ^2D^e_{3/2}$ | 1.082 | 1.088 | 13.563 | 13.570 | 39.703 | 39.700 | 114.319 | 114.266 |
| $4d\ ^2D^e_{5/2}$ | 1.082 | 1.088 | 13.564 | 13.570 | 39.708 | 39.706 | 114.365 | 114.320 |
| $4f\ ^2F^o_{5/2}$ | 1.083 | 1.088 | 13.565 | 13.573 | 39.713 | 39.710 | 114.374 | 114.342 |
| $4f\ ^2F^o_{7/2}$ | 1.083 | 1.088 | 13.566 | 13.573 | 39.715 | 39.712 | 114.397 | 114.379 |

Table II. Binding energies (in Rydbergs) of all $1s^2 2s^2\ ^1S_0^e$ ground state ions in the present work and in NIST data [34].

| Ion | Present | NIST | Ion | Present | NIST |
|---|---|---|---|---|---|
| Be Z=4 | 0.682 | 0.685 | $Si^{+10}$ Z=14 | 35.008 | 35.012 |
| $B^+$ Z=5 | 1.844 | 1.849 | $S^{+12}$ Z=16 | 47.942 | 47.930 |
| $C^{+2}$ Z=6 | 3.514 | 3.520 | $Ar^{+14}$ Z=18 | 62.920 | 62.897 |
| $N^{+3}$ Z=7 | 5.688 | 5.694 | $Ca^{+16}$ Z=20 | 79.957 | 79.900 |
| $O^{+4}$ Z=8 | 8.365 | 8.371 | $Ti^{+18}$ Z=22 | 99.069 | 98.960 |
| $Ne^{+6}$ Z=10 | 15.228 | 15.234 | $Cr^{+20}$ Z=24 | 120.276 | 120.100 |
| $Mg^{+8}$ Z=12 | 24.107 | 24.100 | $Fe^{+22}$ Z=26 | 143.601 | 143.953 |

Table III. Binding energies (in Rydbergs) of all $1s^2 2s 2p\ ^3P_0^o$ metastable ions in the present work and in NIST data [34].

| Ion | Present | NIST | Ion | Present | NIST |
|---|---|---|---|---|---|
| Be Z=4 | 0.482 | 0.485 | $Si^{+10}$ Z=14 | 33.456 | 33.464 |
| $B^+$ Z=5 | 1.502 | 1.509 | $S^{+12}$ Z=16 | 46.122 | 46.115 |
| $C^{+2}$ Z=6 | 3.034 | 3.042 | $Ar^{+14}$ Z=18 | 60.831 | 60.813 |
| $N^{+3}$ Z=7 | 5.072 | 5.082 | $Ca^{+16}$ Z=20 | 77.598 | 77.546 |
| $O^{+4}$ Z=8 | 7.614 | 7.625 | $Ti^{+18}$ Z=22 | 96.436 | 96.334 |
| $Ne^{+6}$ Z=10 | 14.209 | 14.220 | $Cr^{+20}$ Z=24 | 117.365 | 117.202 |
| $Mg^{+8}$ Z=12 | 22.821 | 22.820 | $Fe^{+22}$ Z=26 | 140.411 | 140.780 |

**Figure Captions**

1. $1s^2 2s^2$ $^1S_0^e$ ground state total photoionization cross section for Be, $B^+$, $C^{+2}$, $N^{+3}$, $O^{+4}$, $Ne^{+6}$, $Mg^{+8}$, $Si^{+10}$, $S^{+12}$, $Ar^{+14}$, $Ca^{+16}$ and $Fe^{+22}$ up to the $1s^2 4f$ thresholds of the three-electron final state ion calculated using the relativistic Breit-Pauli R-matrix methodology.

2. As Fig. 1 but for the $1s^2 2s2p$ $^3P^o$ metastable state.  The cross section shown is a statistical average of the three individual $^3P_j^o$ cross sections.

3. Photoionization cross section of ground state Be from 9.2 eV to 13.3 eV; (a) present BPRM result, (b) present nonrelativistic result, (c) experiment [15].  Both theoretical cross sections were calculated with energy step $\Delta E = 68$ µeV, and convoluted with FWHM = 12 meV to match experiment.

4. Photoionization cross section of ground state Be from 16 eV to 21.5 eV; (a) present BPRM result, (b) present nonrelativistic result, (c) experiment [16].  Both theoretical cross sections were calculated with energy step $\Delta E = 68$ µeV, and convoluted with FWHM = 5 meV to match experiment.

5. Photoionization cross section of $B^+$ from 22.5 eV to 31.25 eV.  The theoretical results are a weighted sum of ground state (71%) and metastable $^3P_1^o$ state (29%) cross sections [17]; (a) present BPRM result, (b) present nonrelativistic result, (c) previous BPRM result [17] multiplied by 1.05, (d) experiment [17].  Both present results were calculated with energy step $\Delta E = 13.6$ µeV and convoluted with FWHM = 25 meV to match experiment.  The previous BPRM result was convoluted in the same manner, and the ground and metastable state results were shifted by -22 meV and 4 meV, respectively, to match the measurement.

6. Photoionization cross section of $C^{+2}$ from 41 eV to 57 eV. The theoretical results are a weighted sum of ground state (60%) and metastable $^3P_0^o$ state (30%), $^3P_1^o$ state (5%), and $^3P_2^o$ state (5%) cross sections [18]; (a) present BPRM result, (b) present nonrelativistic result, (c) previous BPRM result [18], (d) experiment [18]. All present results were calculated with energy step $\Delta E = 12.2$ μeV and convoluted with FWHM = 30 meV to match experiment. The previous BPRM result was convoluted in the same manner.

7. Photoionization cross section of $N^{+3}$ from 65 eV to 90 eV. The theoretical results are a weighted sum of ground state (65%) and metastable state (35%) cross sections [19]; it was assumed that the metastable fractions were statistical. Shown are (a) present BPRM result, (b) present nonrelativistic result, and (c) experiment [19]. Both present results were calculated with energy step $\Delta E = 13.6$ μeV and convoluted with FWHM = 230 meV to match experiment.

8. Photoionization cross section of $O^{+4}$ from 95 eV to 130 eV. The theoretical results are a weighted sum of ground state (50%) and metastable state (50%) cross sections [21]; it was assumed that the metastable fractions were statistical. Shown are (a) present BPRM result, (b) present nonrelativistic result, and (c) experiment [21]. Both present results were calculated with energy step $\Delta E = 13.6$ μeV and convoluted with FWHM = 250 meV to match experiment.

9. Comparison of the present BP photoionization cross sections, (a) ground state, (c) metastable state for Be with OP results, (b) ground state, (d) metastable state [22]. The metastable BP results shown are a statistical average of the three $^3P^o_j$ cross sections.

10. As Fig. 9 for $Ne^{+6}$.

11. As Fig. 9 for $Ar^{+14}$.

12. As Fig. 9 for $Fe^{+22}$.

13. Total photoionization cross section of Be; (a) present BP ground state result, (b) VRM ground state result for [23], (c) present BP metastable state result, (d) VRM metastable state result [24]. The BP metastable state cross section is a statistical average of the three $^3P^o_j$ cross sections.

14. As Fig. 13 for B$^+$; the VRM results are from Refs. 25 and 26 for ground and metastable states, respectively.

15. As Fig. 13 for C$^{+2}$; the VRM results are from Refs. 27 and 28 for ground and metastable states, respectively.

16. Total photoionization cross section of C$^{+2}$; (a) present BP ground state result, (b) ground state result of Ref. 30, (c) present BP metastable state result, (d) metastable state result of Ref. 30. The BP metastable state cross section is a statistical average of the three $^3P_j^o$ cross sections.

17. As Fig. 16 for N$^{+3}$.

18. As Fig. 16 for O$^{+4}$, but compared with results from Ref. 31.

19. As Fig. 16 but compared with the relativistic results of Ref. 29.

20. (color online) Photoionization cross sections of Ne$^{+6}$ for (a) ground state, and (b) metastable state photoionization. The BP calculations are shown by solid lines, and the *LS* calculations are shown by dashed lines. The BP metastable state cross section is a statistical average of the three $^3P_j^o$ cross sections.

21, As Fig. 20 for S+$^{12}$.

22. As Fig. 20 for Fe$^{+22}$.

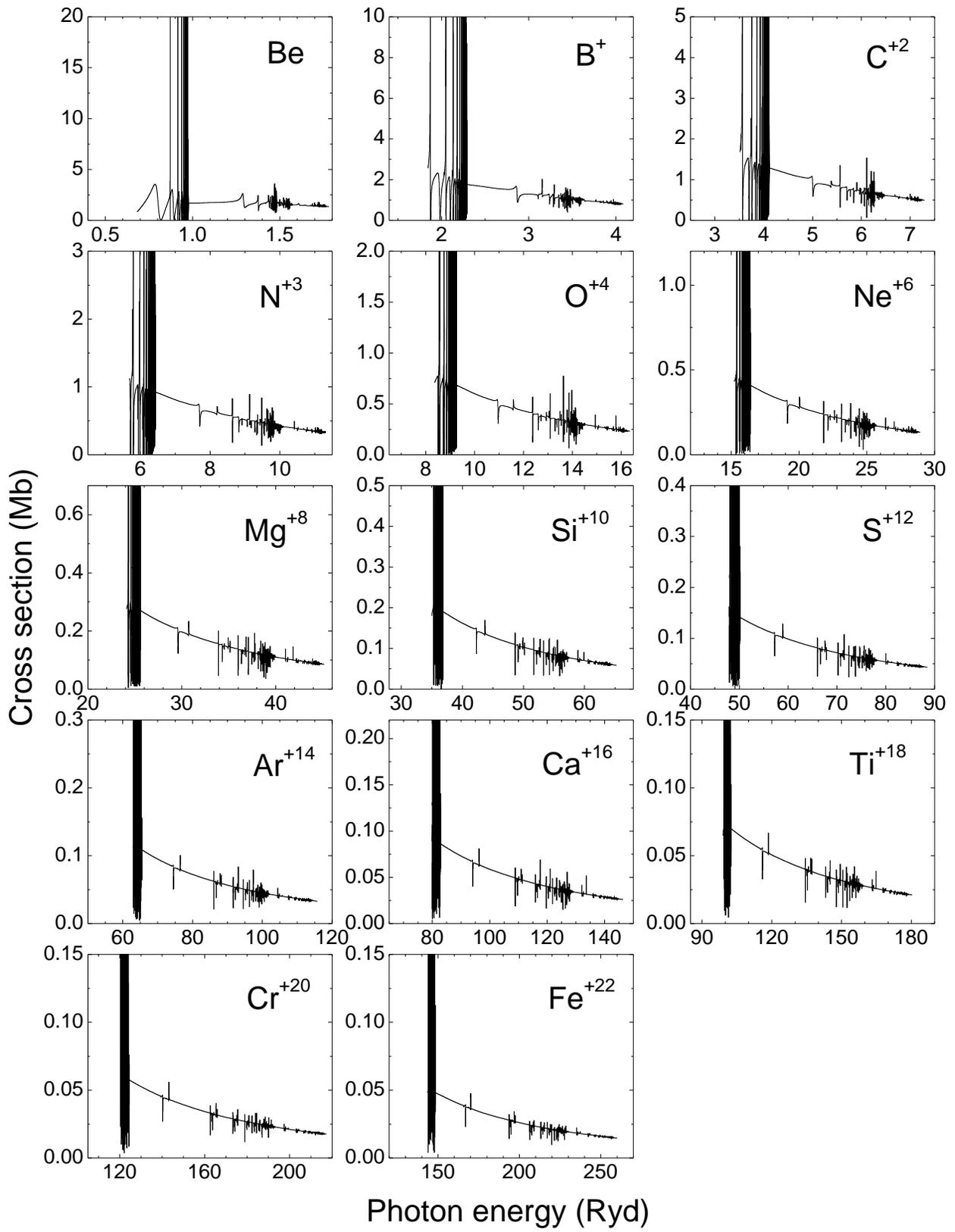

Fig.1

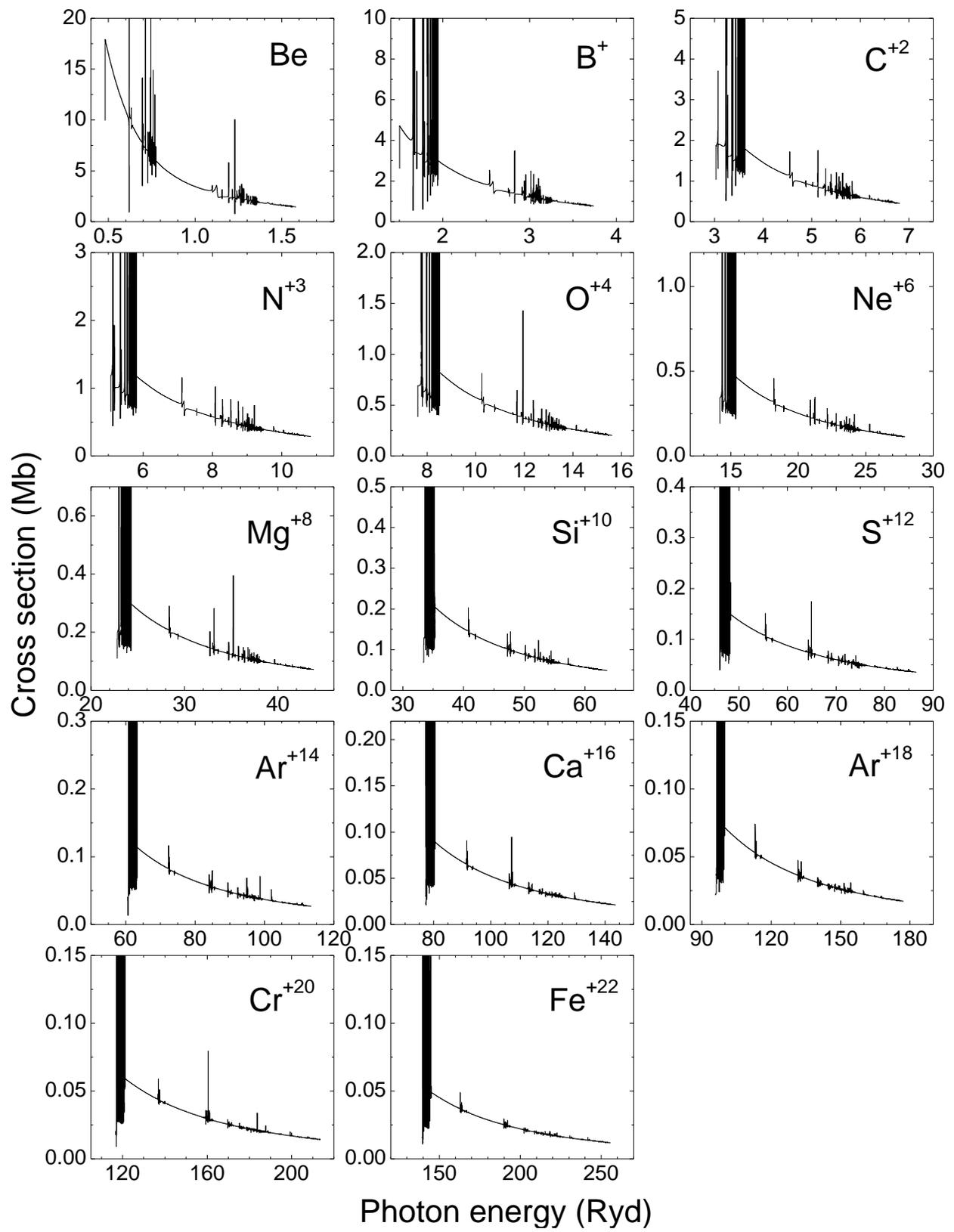

Fig. 2

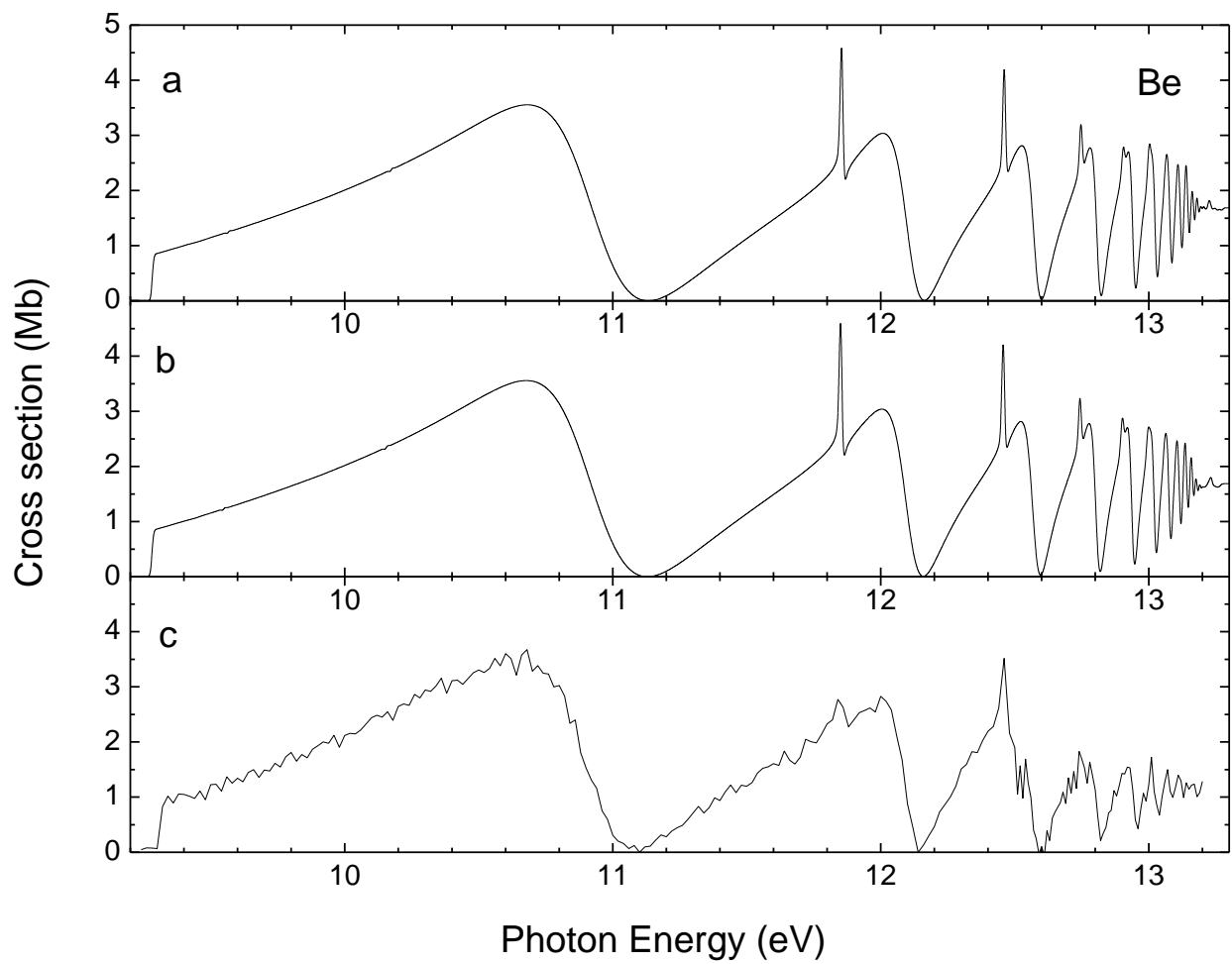

Fig. 3

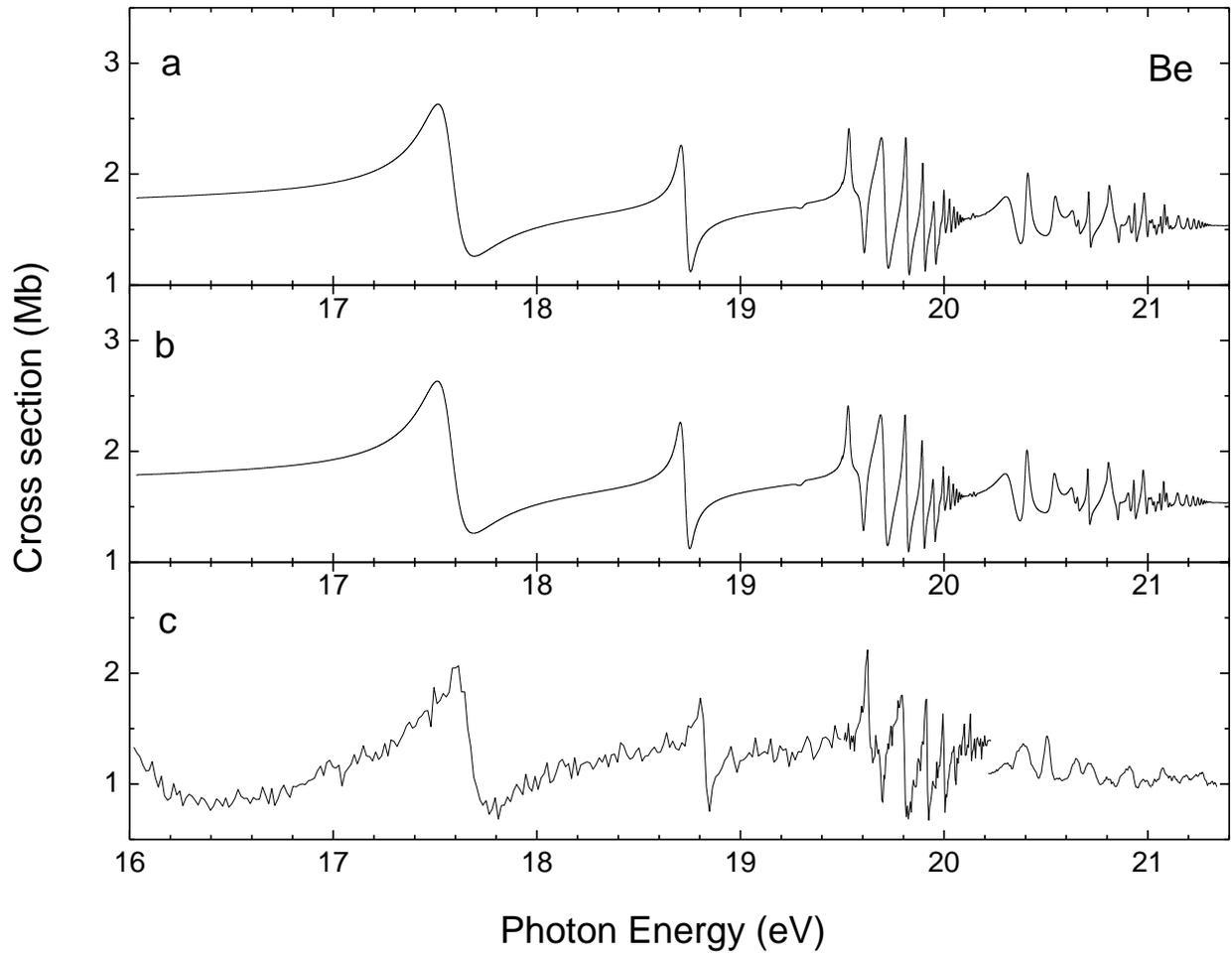

Fig. 4

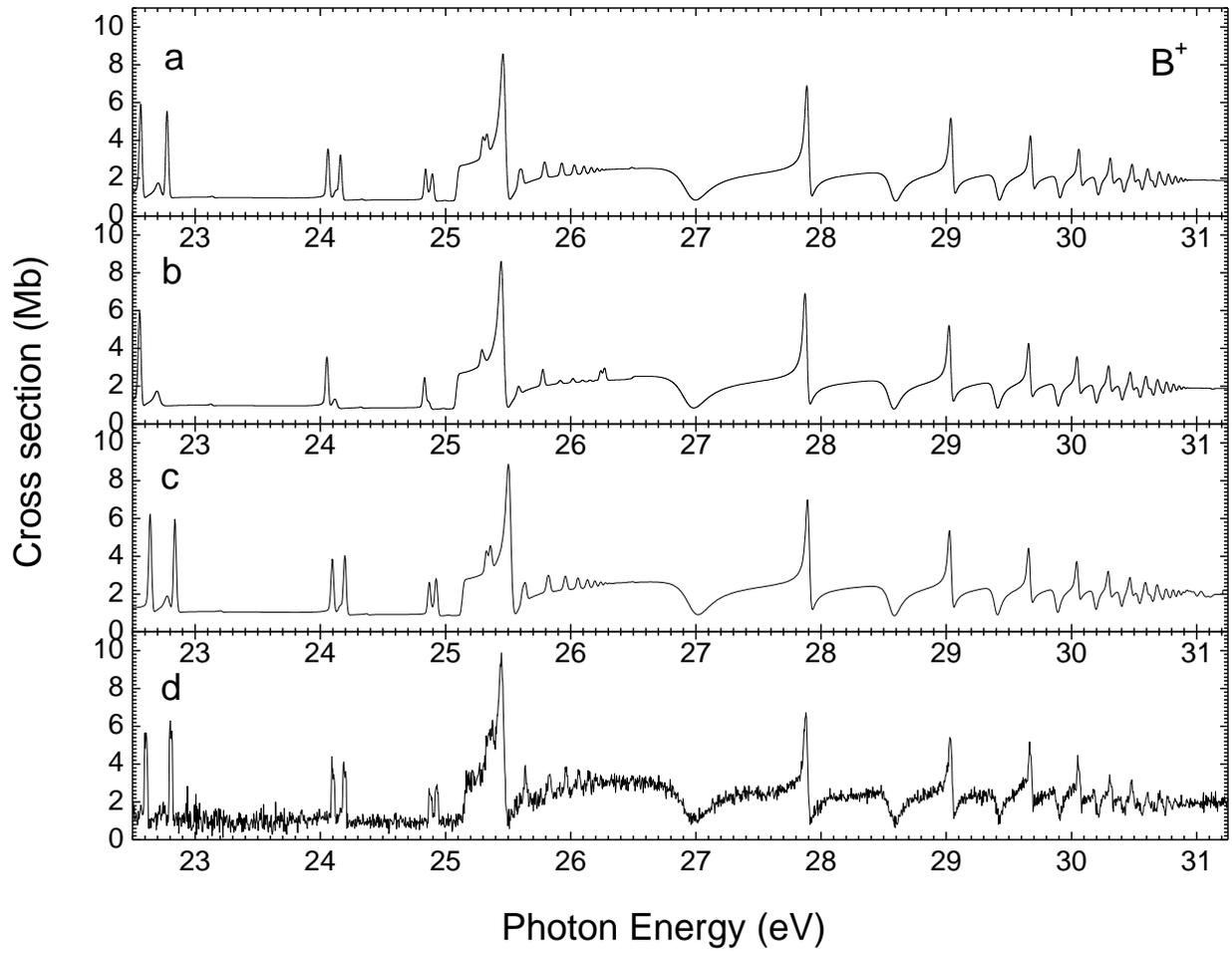

Fig. 5

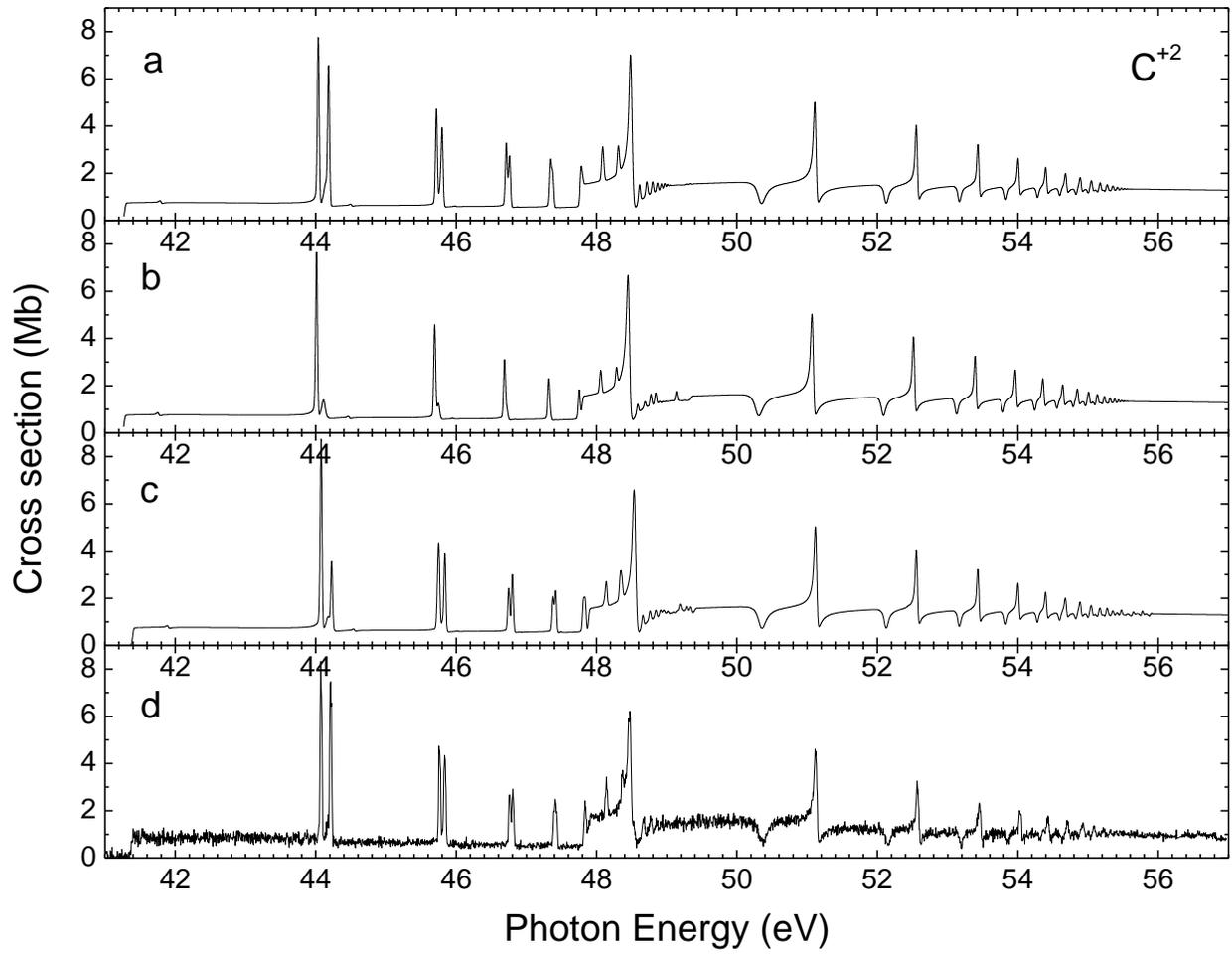

Fig. 6

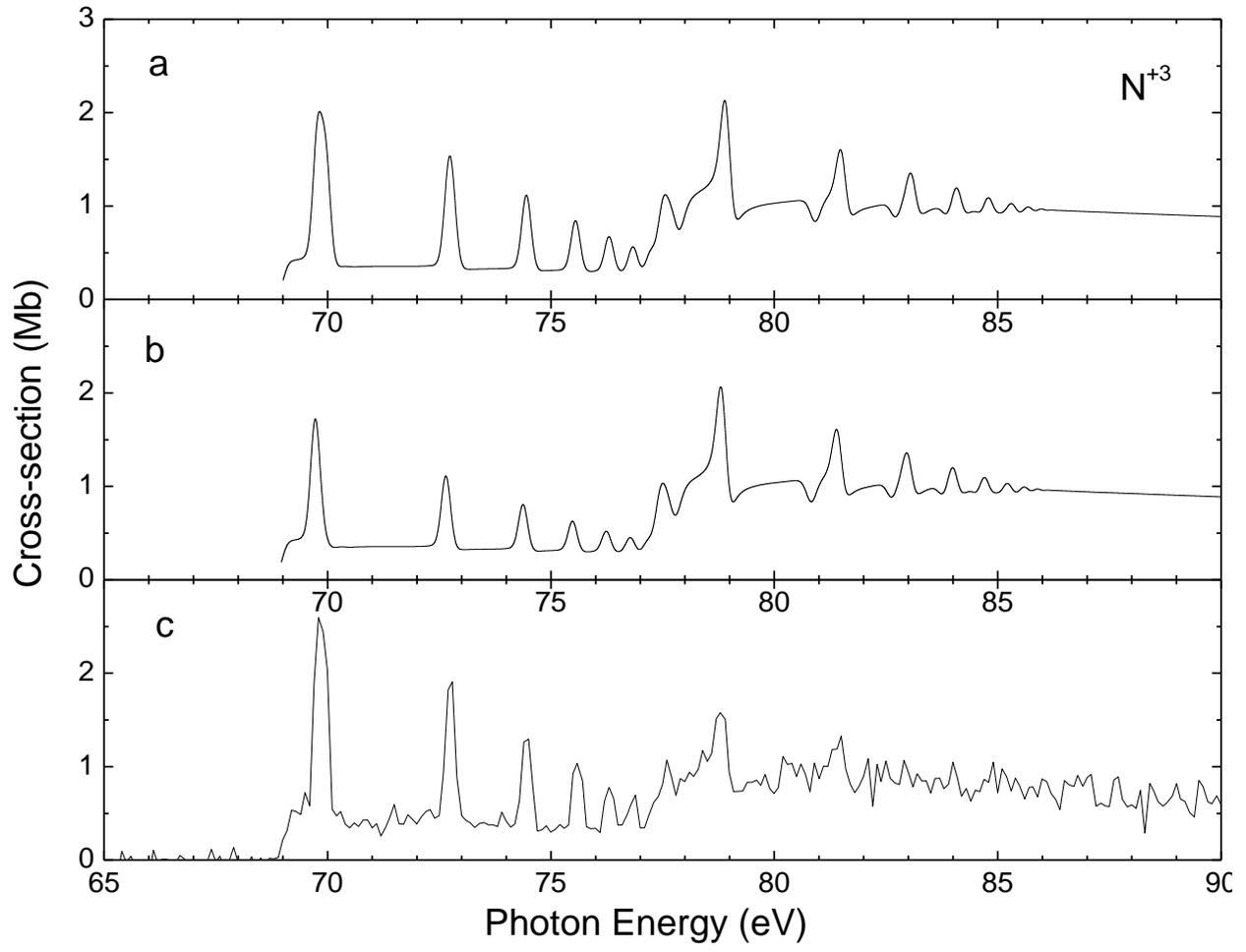

Fig. 7

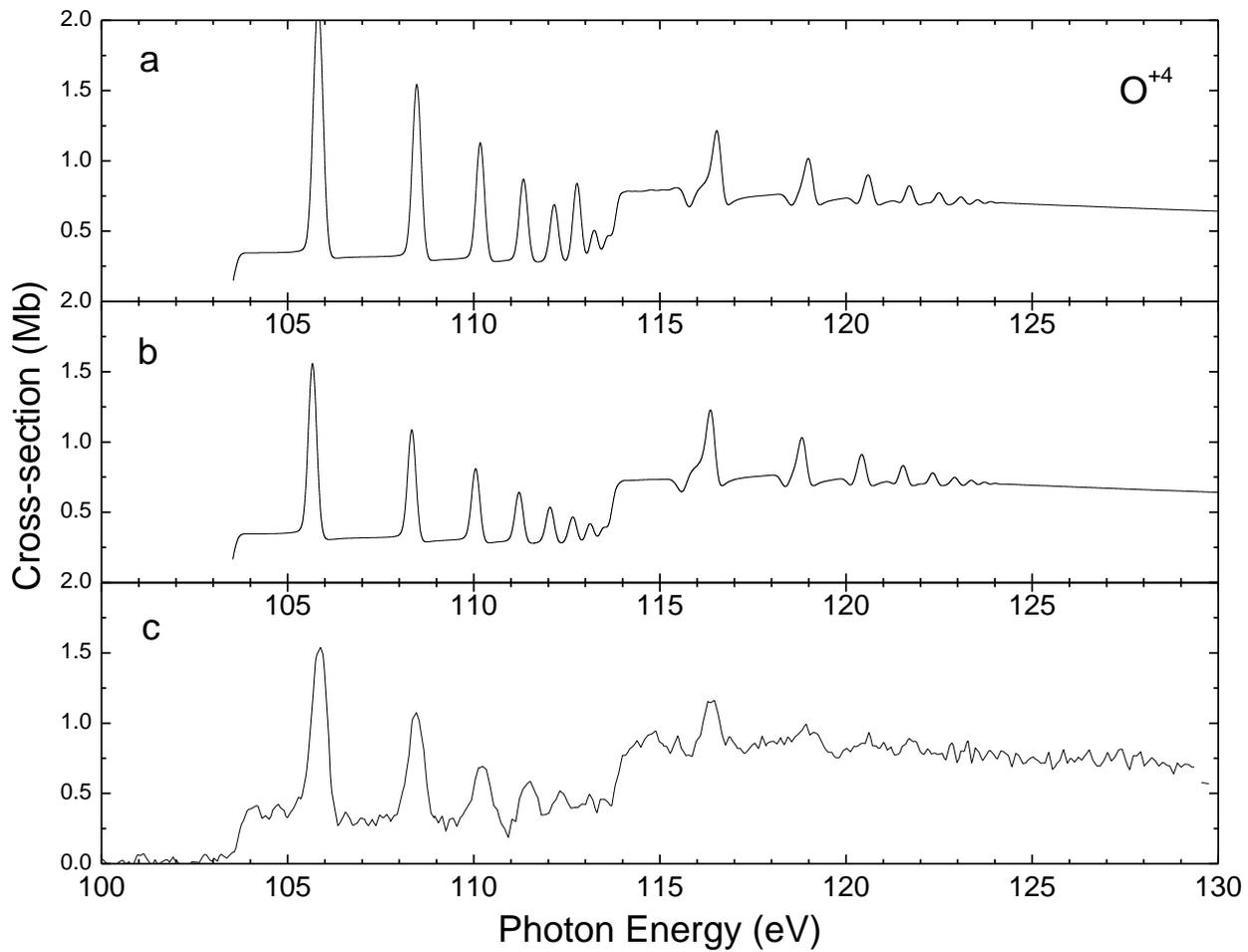

Fig. 8

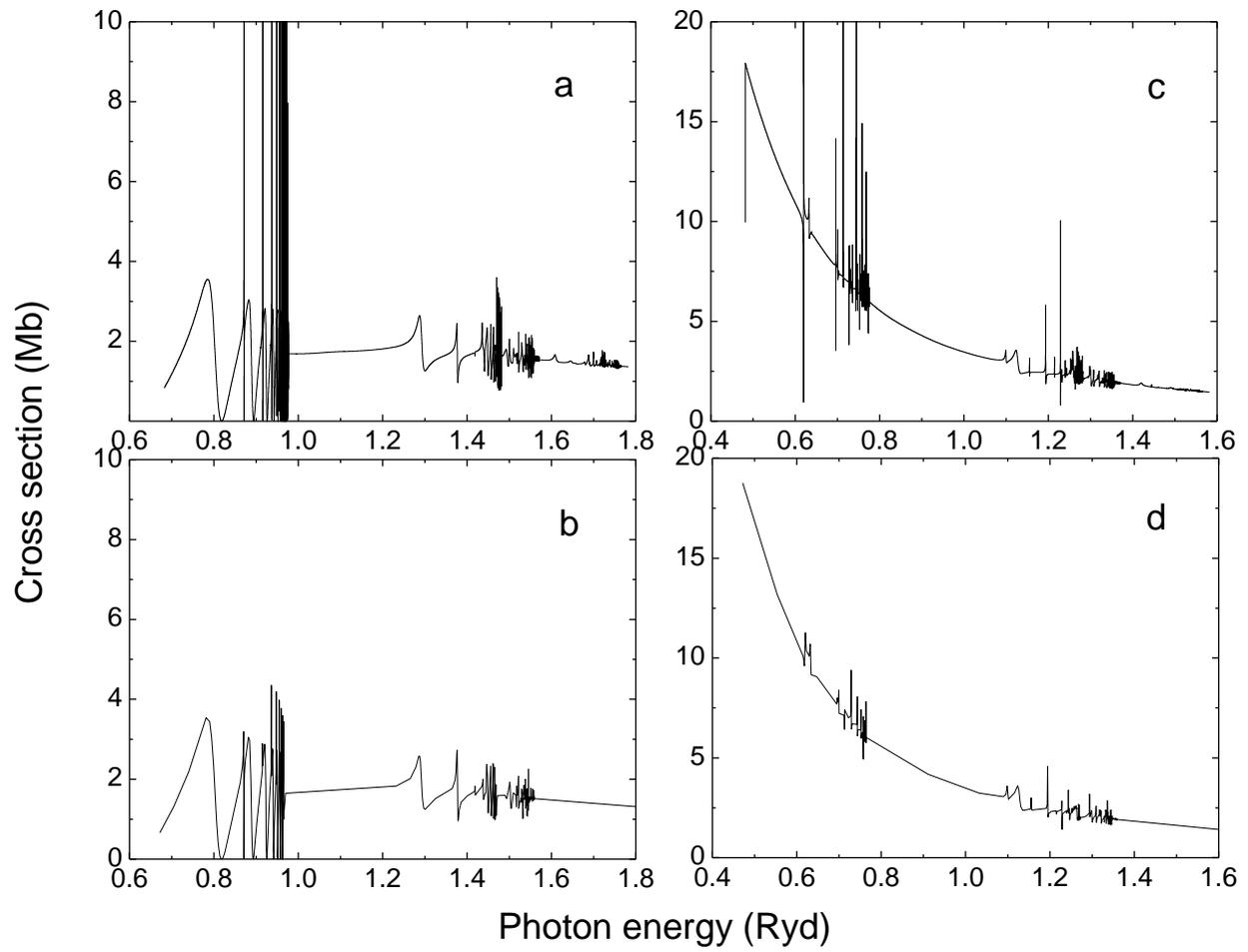

Fig. 9

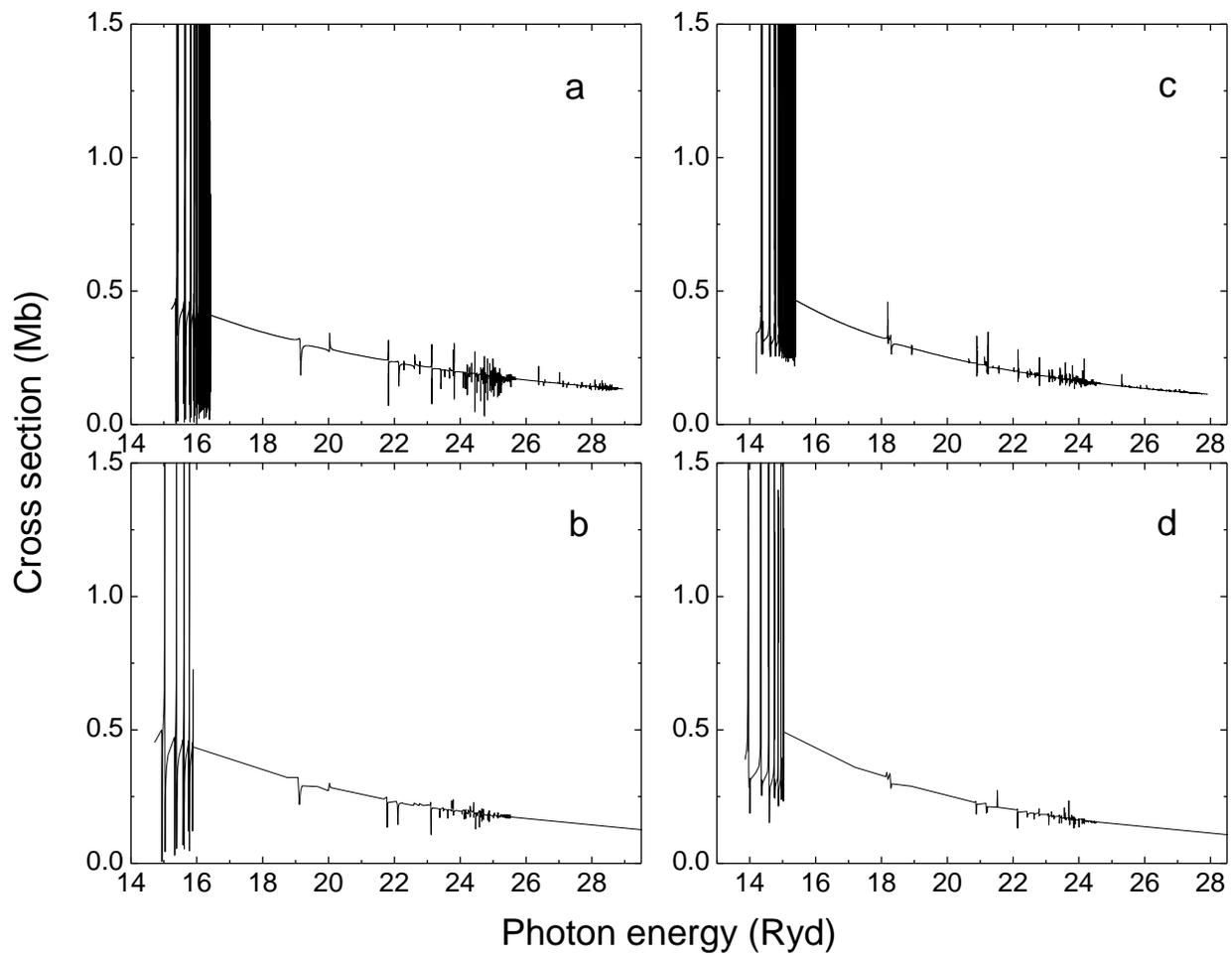

Fig. 10

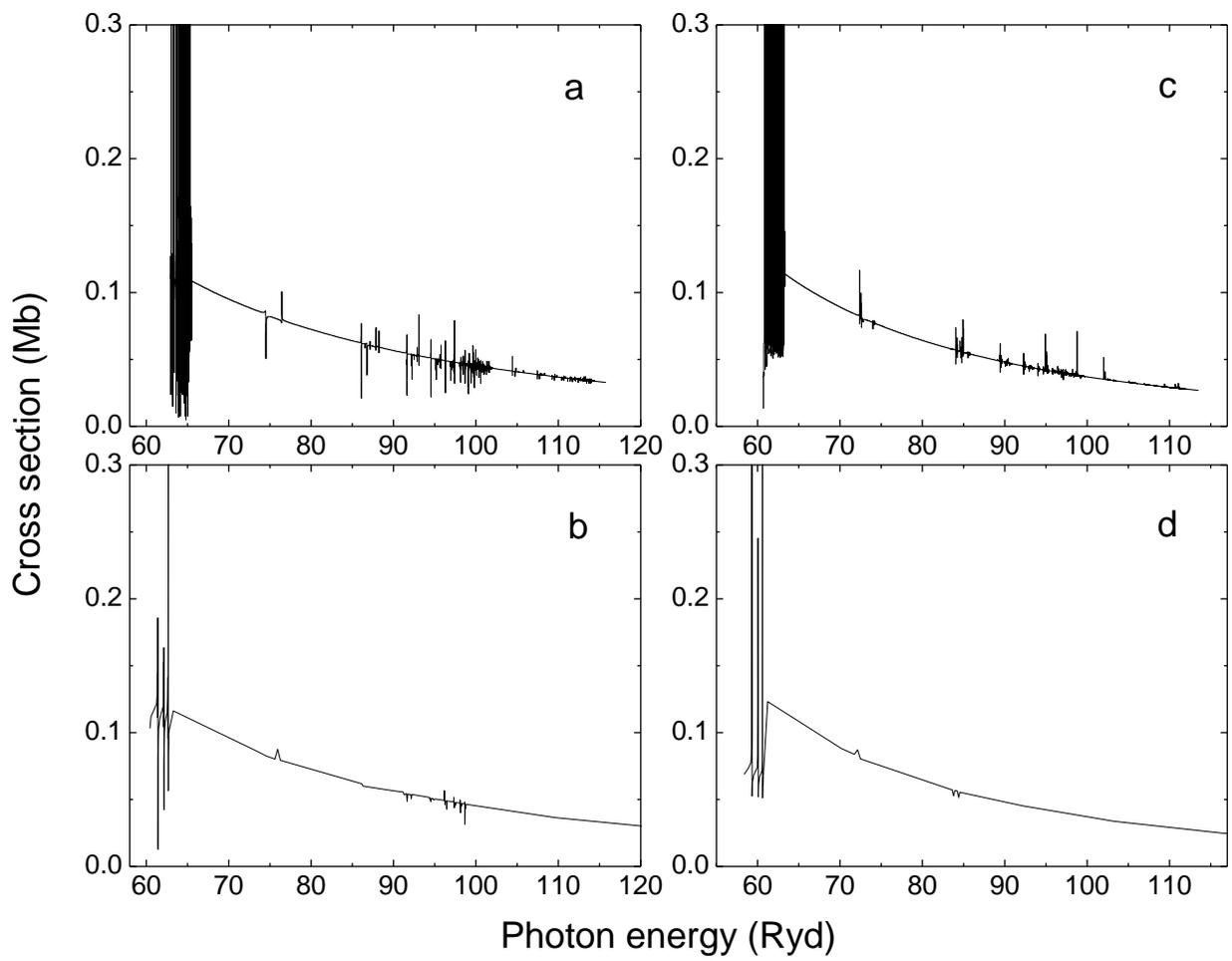

Fig. 11

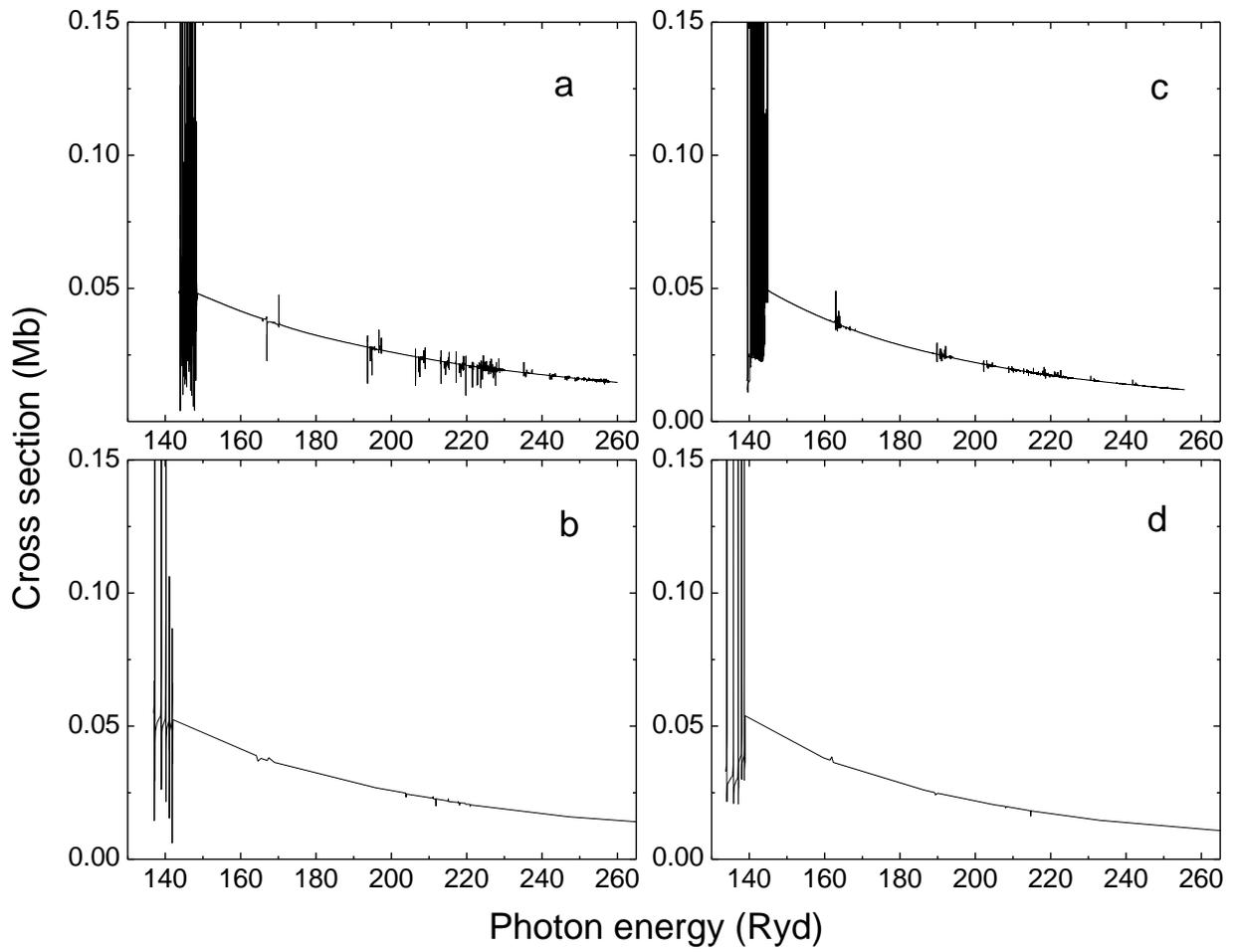

Fig. 12

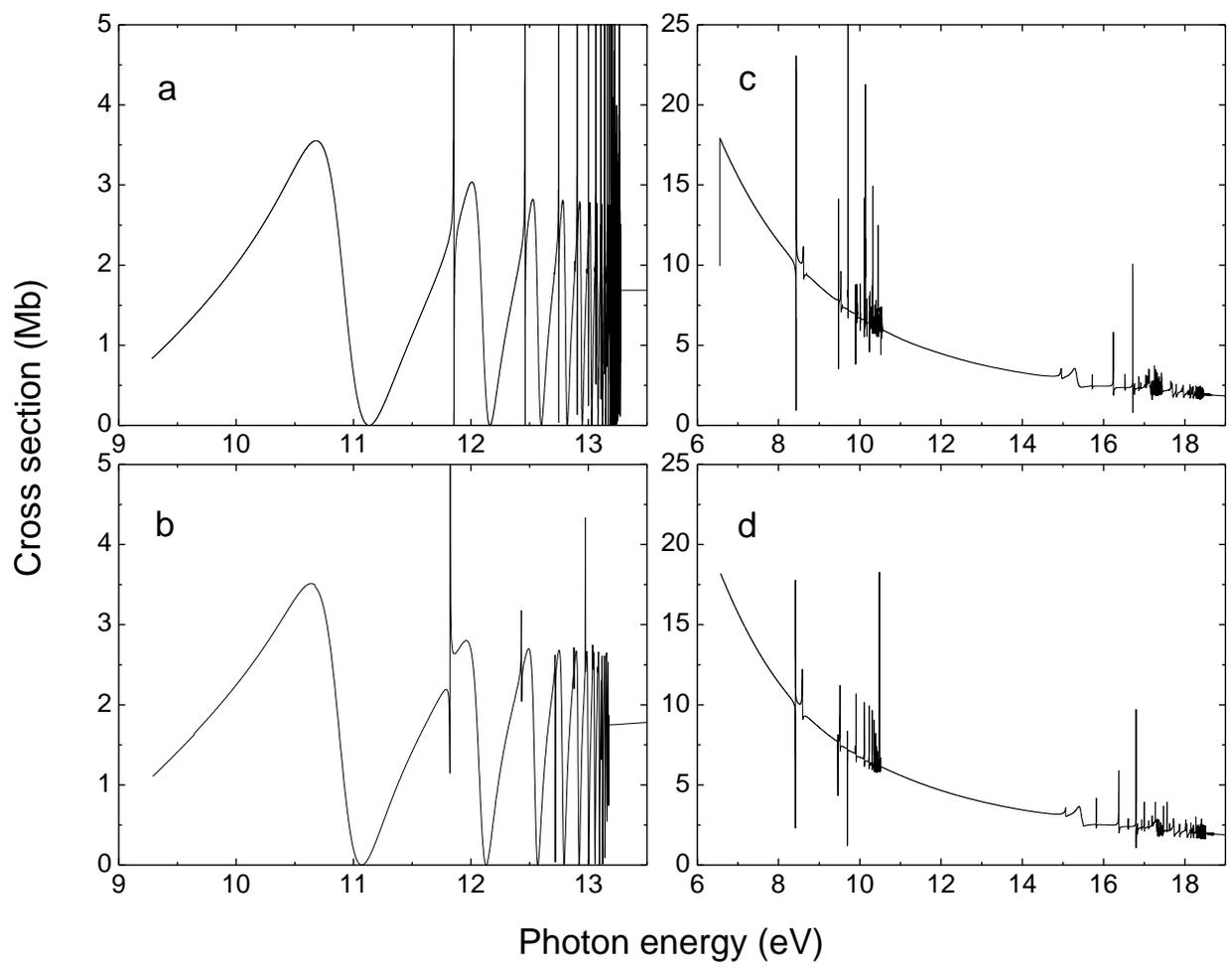

Fig. 13

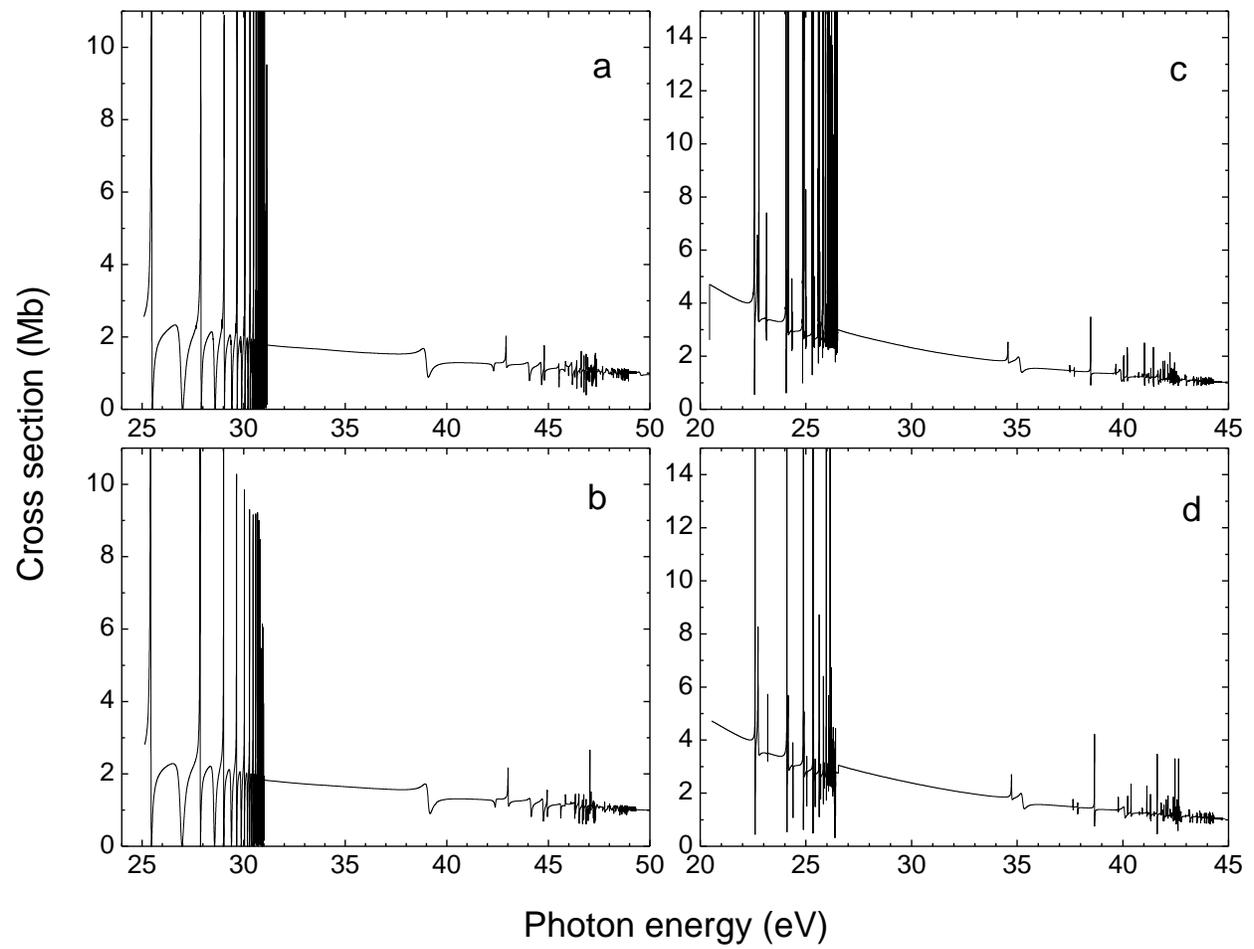

Fig. 14

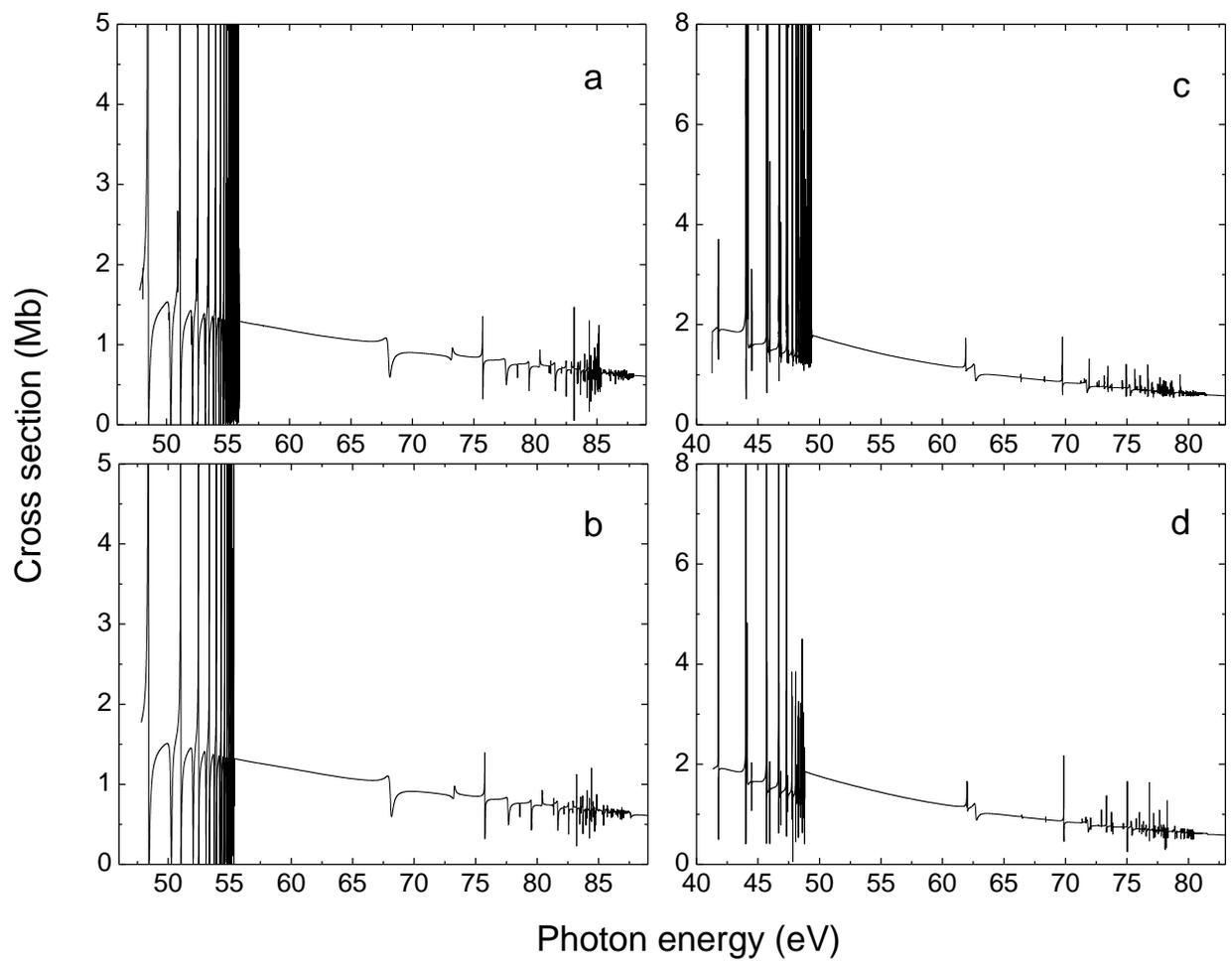

Fig. 15

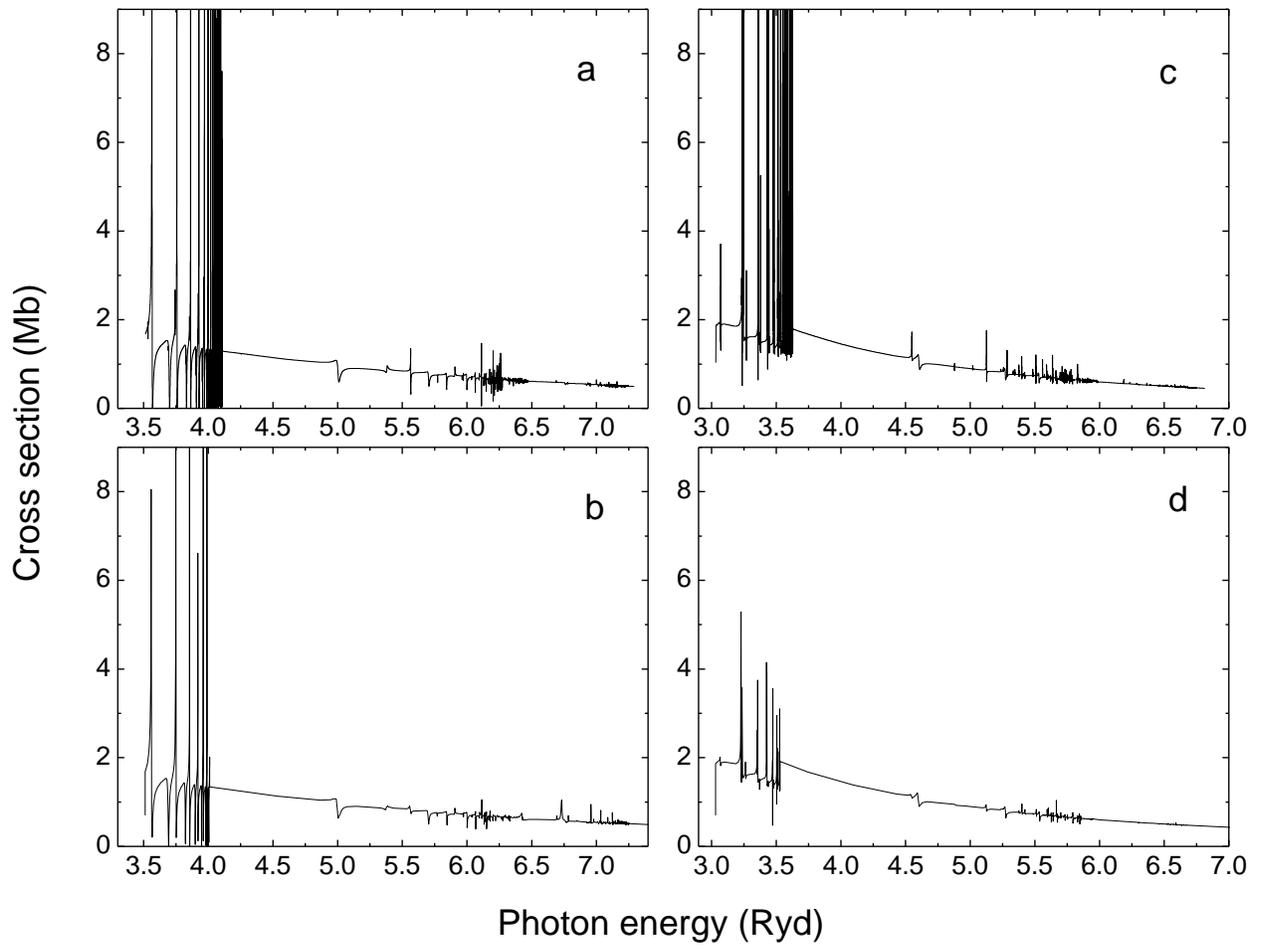

Fig. 16

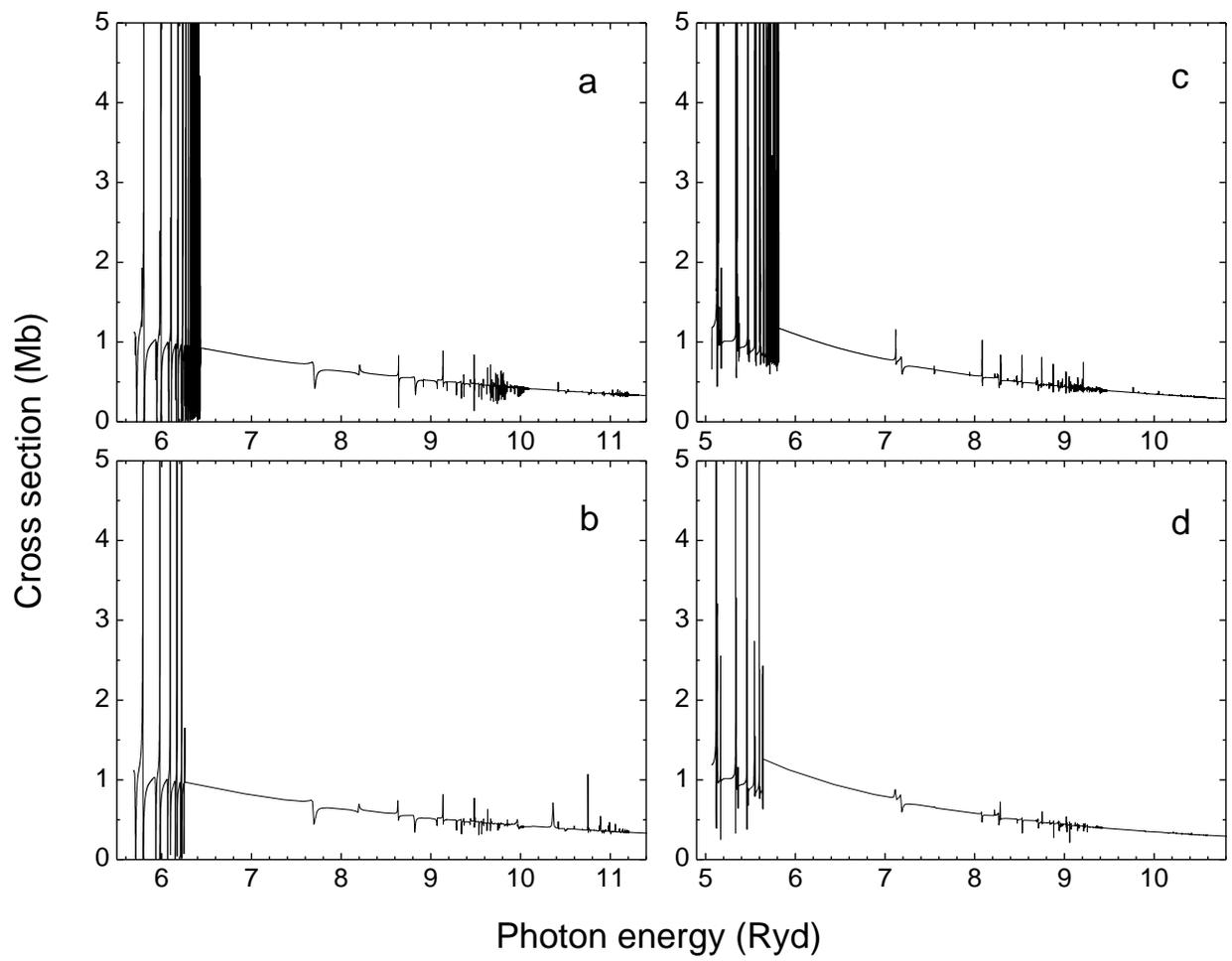

Fig. 17

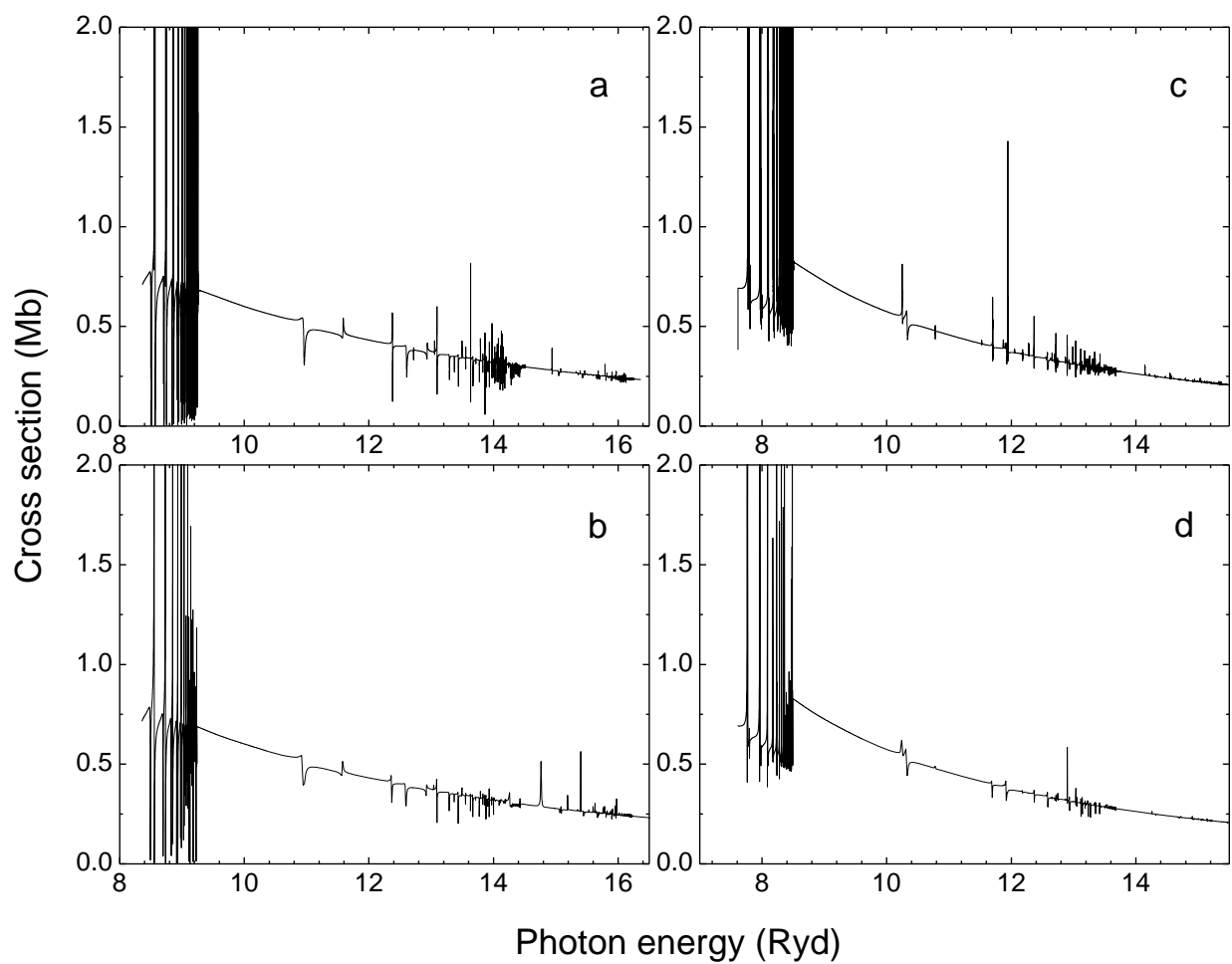

Fig. 18

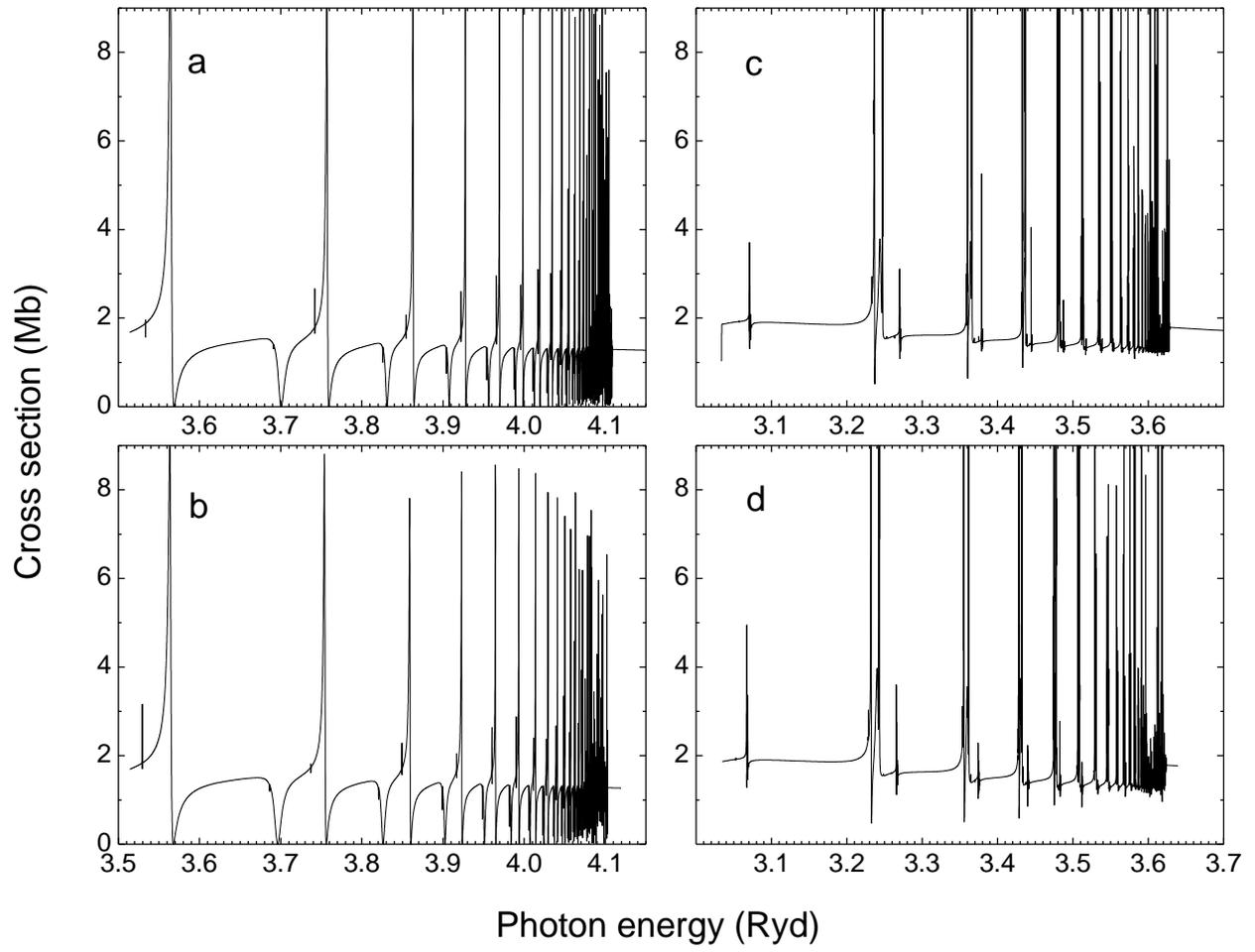

Fig. 19

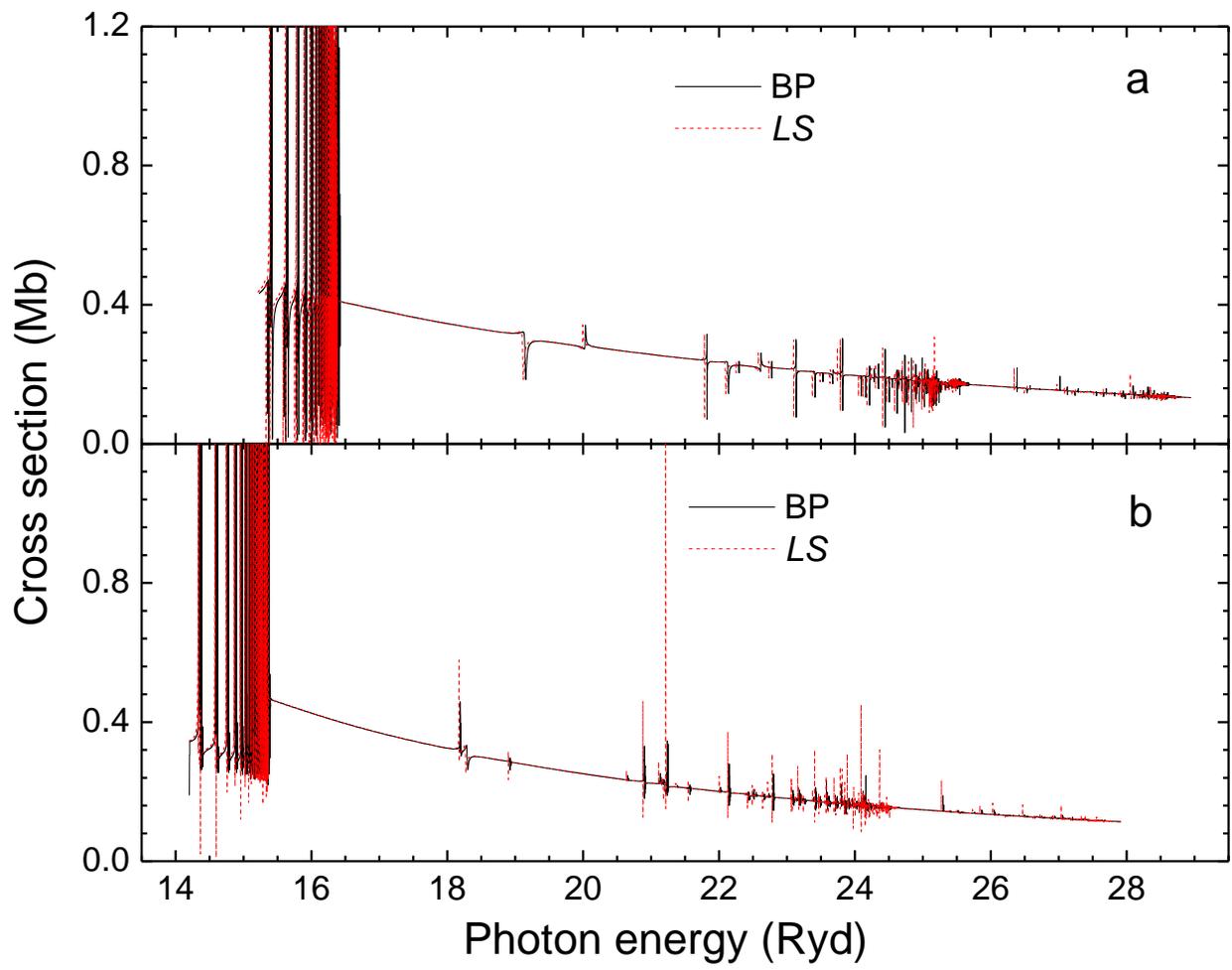

Fig. 20

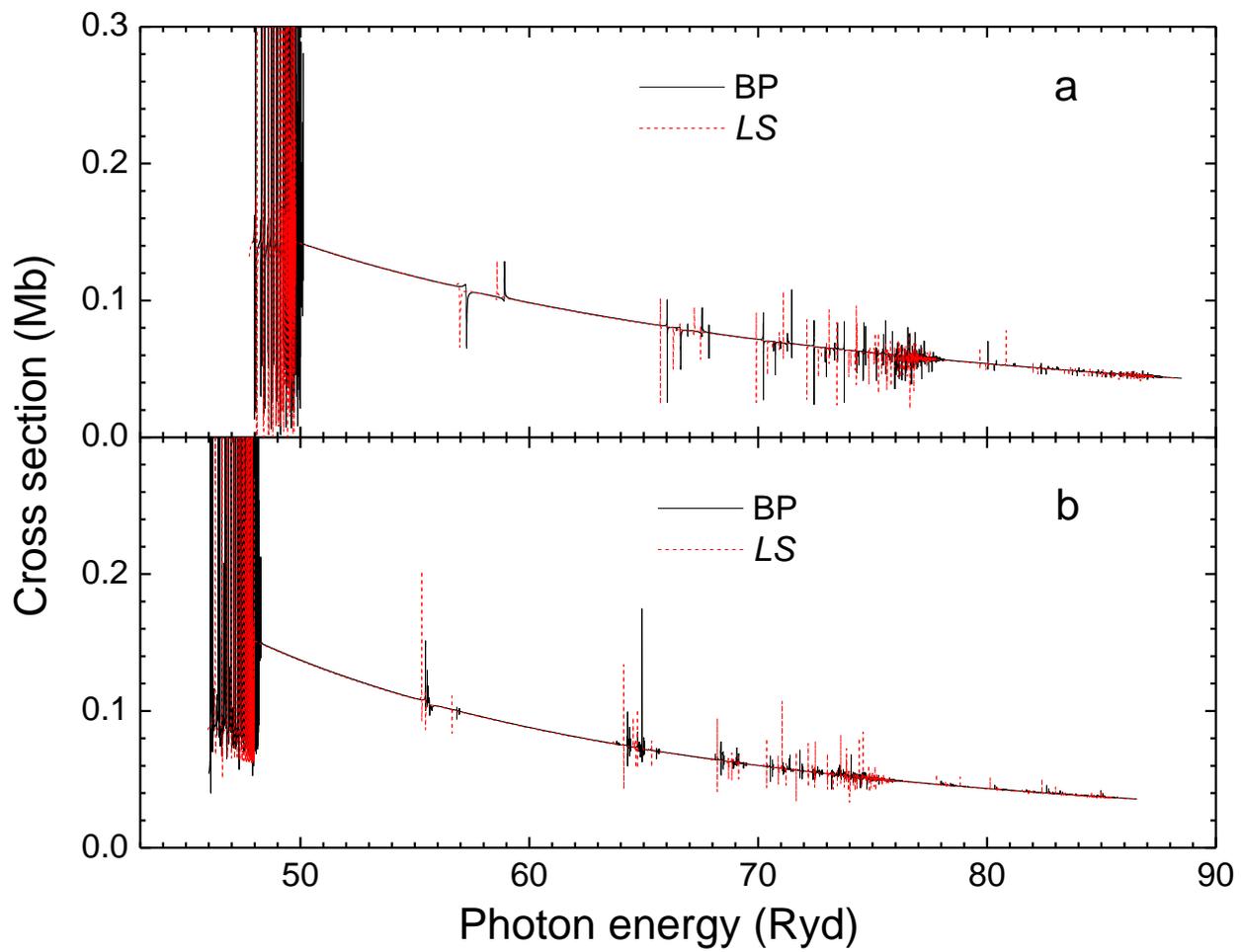

Fig. 21

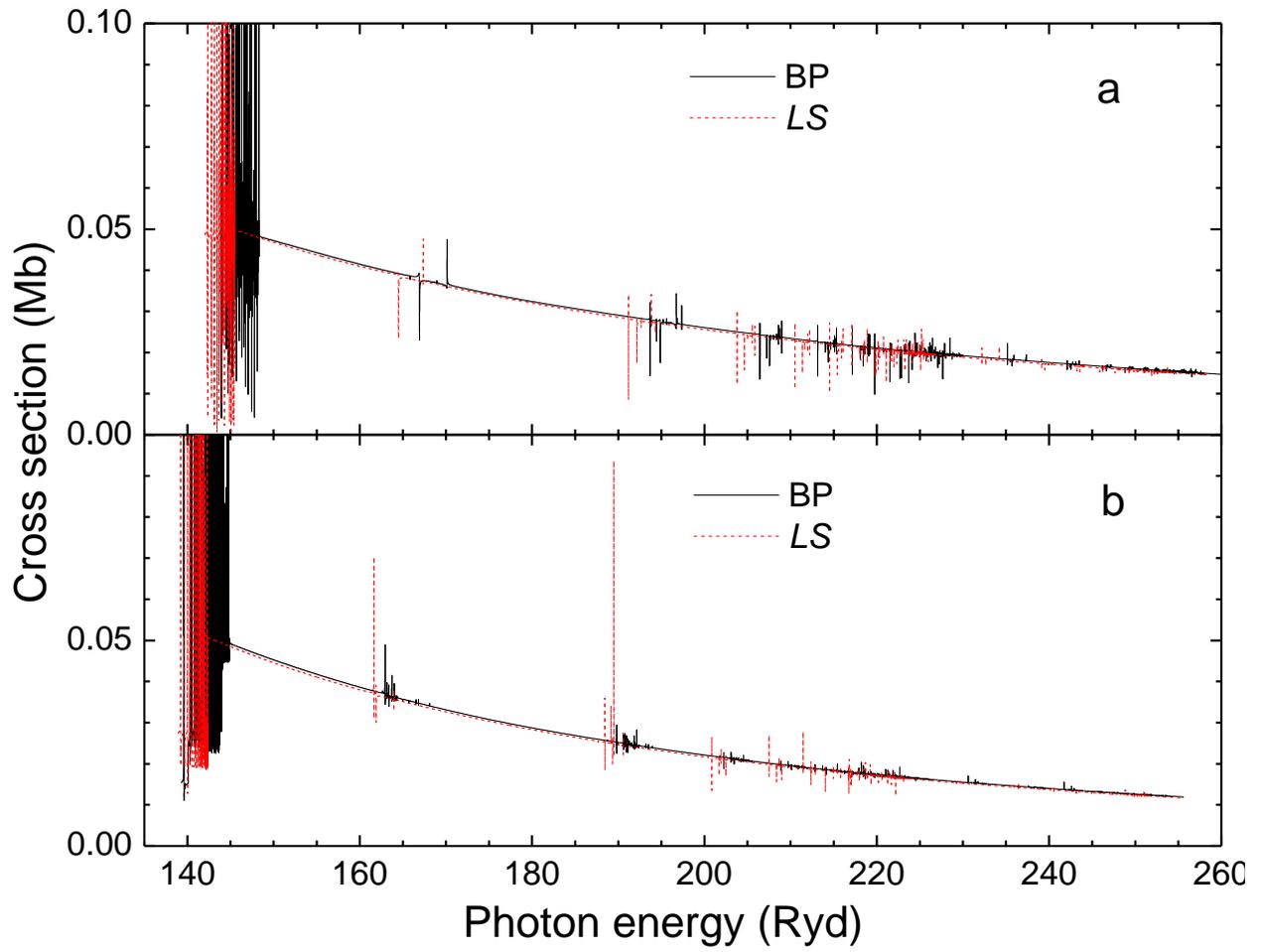

Fig. 22